\documentclass[fp,twocolumn]{jpsj_mod}
\usepackage{latexsym}
\usepackage{amsmath}
\usepackage{amssymb}
\usepackage{bm}
\usepackage{wasysym}
\usepackage{widetext}
\usepackage{graphicx}
\newcommand{\hide}[1]{}
\newcommand{\veps}{\varepsilon}

\def\bfr{{\bf r}}
\def\ra{\rangle}
\def\la{\langle}

\def\veps{\varepsilon}
\def\ua{\uparrow}
\def\da{\downarrow}

\usepackage{tikz}
\usetikzlibrary{matrix}
\usetikzlibrary{arrows}

\graphicspath{{./figures/}}
\usepackage{float}

\newcommand{\bolOmega}{{\bf m}}

\newcommand{\be}{\mathrm{e}}
\newcommand{\aclockwise}{\raisebox{-0.3ex}{\rotatebox{90}{$\circlearrowleft$}}}
\begin{document}
\title{Non-Abelian Aharonov-Casher Phase Factor in Mesoscopic Systems}
\author{Yshai Avishai$^{1,2,3}$\thanks{yshai@bgu.ac.il}, Keisuke Totsuka$^2$, and Naoto Nagaosa$^{4,5}$}
\inst{$^1$ {\small Department of Physics, Ben-Gurion University, Beer-Sheva 84105, Israel} \\
$^2$ {\small Yukawa Institute for Theoretical Physics, Kyoto University, 
Kyoto 606-8502, Japan} \\
$^3$ {\small Department of Physics, New-York University at Shanghai, China} \\
$^4$ {\small Department of Applied Physics, The University of Tokyo, Bunkyo, Tokyo, 113-8656, Japan} \\
$^5${\small RIKEN Center for Emergent Matter Science (CEMS), Wako, Saitama, 351-0198, Japan}}
\abst{%
The matrix-valued {\it Aharonov-Casher phase  factor}  $F_{\text{AC}}$ 
(related to the c-number {\it Aharonov-Casher phase} $\lambda_{\text{AC}}$) plays an important role in the physics 
of mesoscopic systems in which spin-orbit coupling is relevant.  
Yet, its relation to experimental observables is rather elusive. 
Based on the SU(2)-gauge-invariant formulation of the Schr\"odinger equation, we relate $F_{\text{AC}}$ to measurable quantities 
in electronic interferometers subject to electric fields that generate Rashba or Dresselhaus spin-orbit coupling.  
Specifically, we consider electron transmission through (i) a single-channel ring interferometer 
 and (ii) a two-channel square interferometer. 
In both examples, we derive the closed expressions of the conductance and 
show them to be simple rational functions of the {\it traceful part} of $F_{\text{AC}}$.  
In the second case, we also derive a closed expression for the electron spin polarization vector and find it to be a simple function 
of {\it both} the traceful and traceless parts of $F_{\text{AC}}$. 
This analysis then suggests a direct way for an experimental access to this elusive quantity.
}
\maketitle
\newpage
\section{Introduction}
The Aharonov-Casher phase factor (ACPF, denoted hereafter as $F_{\text {AC}}$)\cite{Aharonov-C-84} is the SU(2) analog 
of the U(1) Aharonov-Bohm {\em phase factor} $F_{\text {AB}}=\be^{\frac{i e}{\hbar c} \oint {\bf A} \cdot d {\bf s}}\equiv \be^{ i \phi_{\text{AB}}}$, 
 wherein the electromagnetic vector potential ${\bf A}$ is integrated along a non-contractible loop and  $\phi_{\text{AB}}$ 
 is the Aharonov-Bohm {\em phase}. 
 On the other hand, $F_{\text {AC}}$ is related to (albeit richer than) the somewhat more familiar quantity, the Aharonov-Casher phase   
(ACP, denoted as  $\lambda_{\text{AC}}$). 
The crucial r\^ole  played by (non-integrable) phase factors in gauge-field theories has been emphasized in Ref.~\citen{Wu-Yang},
wherein the subtle distinction between the physical meanings of phase and phase factor is clarified.  
In the Aharonov-Casher effect, one studies the effect of an electric field on the dynamics 
of a quantum-mechanical particle with spin $s$ and a magnetic moment 
$\mu$. Due to relativistic dynamics, the particle feels an effective magnetic field (in its rest frame), that interacts with its magnetic moment.  
In the Pauli equation for the electron, this scenario is accounted for by the spin-orbit coupling (SOC).  
In a paradigmatic example of demonstrating the Aharonov-Casher effect, a particle is 
confined to the $x$-$y$ plane threaded by a perpendicular charged wire 
with constant longitudinal charge density $\eta$ that produces a radial electric field. 
For spin $s=\frac{1}{2}$, upon completing a motion along a closed planar curve 
around the wire, the spin-up (spin-down) component of the particle will acquire a 
phase $2 \pi \beta$ 
($-2 \pi \beta$). This implies that the two-component spinor acquires an SU(2) non-integrable phase factor (ACPF) 
$\binom{\be^{2 \pi i \beta} \ \ \  0}{0 \ \ \ \be^{-2 \pi i \beta}}$ where $\beta=\eta/\eta_0$, with $\eta_0=2 h c/g \mu$.   
Generically however, as we shall see below, when the electric field vector is not 
parallel to the $x$-$y$ plane, the ACPF is not diagonal. 
 
In the present work, we focus on the relevance of the
 ACP/ACPF to the physics of mesoscopic systems, for which electrons are 
 sometimes subject to a strong SOC.  
It will be demonstrated that these (somewhat abstract) 
quantities are closely related to physical observables and hence can be measured and provide useful information 
on the pertinent SOC mechanism inside matter. 
Specifically, we consider electron transmission through various interferometers 
within which SOC is active.
Pertinent experimental and theoretical studies focus mainly 
on the single number ACP (that is, $\lambda_{\mathrm{AC}}$)\cite{Mathur-H-92, Loss-G-B-90,Stern-92,Nitta-M-T-99,Nitta-B-07,Meijer-M-K-02,Frustaglia-R-04,%
Molnar-P-V-04,Borunda-L-K-L-J-S-08, Wang-V-05}.  
Here we shall go further and extend our discussion to elucidate the r\^ole of the matrix-valued phase factor ACPF.  
It is demonstrated that the ACPF provides information on the spin physics much beyond that provided by the ACP.   
Relevant experimental observables related to electron interferometry are the (dimensionless) conductance $g$,   
as well as the transmitted (T) and reflected (R) spin polarization vectors ${\bf P}^{\text{T}}$ and ${\bf P}^{\text{R}}$.  
Hence, the central question addressed in this work is 
how $g$ and ${\bf P}^{\text{T/R}}$ depend on $\lambda_{\mathrm {AC}}$ or, more generally, on the matrix $F_{\text{AC}}$ itself.

In the study of electron 
transport through an interferometer wherein electric  fields are active and generate SOC, 
an important r\^ole is played by time-reversal symmetry that is (non-trivially) respected in the absence of an external magnetic field.  
  Recall that the electron interferometer 
 is the key experimental tool for studying the 
 Aharonov-Bohm (AB) effect in mesoscopic systems 
 subject to weak {\em magnetic} field (wherein time reversal symmetry is explicitly broken) \cite{Imry-book-02}. 
 As a consequence of U(1) gauge invariance,
  the charge conductance $g$ depends solely on the AB phase $\phi_{\text{AB}}$, no matter how it is generated. 
 In contrast to the AB phase factor $F_{\text{AB}} \equiv e^{ i \phi_{\text{AB}}}$, 
 that is a unimodular number, the ACPF,  $F_{\text{AC}}$, is an SU(2) matrix 
 that is defined as a path-ordered (Wilson) integral of an SU(2) vector potential \cite{Anandan-89,Qian-S-94}.  
Thus, strictly speaking, the ACPF is not fully characterized by a single number; it is specified by both a unit vector and an angle.    
However, the {\it single number} $\cos \lambda_{\text{AC}}\equiv \frac{1}{2}\mbox{Tr}F_{\text{AC}}$, which is a gauge-invariant quantity,  
is shown below to have a clear physical content and may be identified with the ACP in the most general situations.      
Moreover, in the special case of a diagonal ACPF, such as $F_{\text{AC}}=e^{i \lambda_{\text{AC}} \sigma_z}$, the components 
of an electron spinor $\binom {\psi_\ua(\theta)}{\psi_\da(\theta)}$ moving adiabatically on a circular 1D ring 
 (parametrized by an azimuthal angle $0 \le \theta \le 2 \pi$), gain respective phases $\pm \lambda_{\text{AC}}$  
(see the discussion in Sec.~\ref{Subsec-ACPF-ACP}), thereby reproducing the ACP in the standard treatment.\cite{Aharonov-C-84} 
  
The presentation in this work is organized as follows.  
Starting from the SU(2)-invariant formulation of the Schr\"odinger equation\cite{Anandan-89,Frohlich-S-93}, 
we first elaborate, in Sec.~\ref{Sec2},  on the definition and evaluation of 
the SU(2) non-integrable phase factors based on the concept of path-ordered integration. 
Being a $2\times2$ SU(2) matrix, the ACPF consists of both the traceful part (that equals $\cos \lambda_{\text{AC}} \mathbf{1}_{2 \times 2}$) 
and the traceless part that has a vectorial property.  
In Sec.~\ref{sec:Ring}, we apply these concepts and techniques to  elucidate the physics of electron transmission through 
a planar 1D ring interferometer under a homogeneous perpendicular 
electric field that leads to a Rashba SOC specified by a position-independent dimensionless strength parameter $\beta$. 
First, we derive the explicit expressions for the ACPF.
Then, the corresponding scattering problem is solved and a closed expression is obtained for the dimensionless conductance 
$g(k,\mathcal{X} ; \beta)$ (here $k$ is the wave number of the incoming electron and $\mathcal{X}$ specifies the 
geometry of the interferometer) that can be written as a simple (rational) function 
$\mathcal{G}_{\bigcirc} (\cos \lambda_{\text{AC}}(\beta); k,\mathcal{X})$ of $\cos \lambda_{\text{AC}}(\beta)$.  
Consequently, it means that (half) the trace $\cos \lambda_{\text{AC}}(\beta)$ of the ACPF, 
though it seems to be rather elusive, can, in fact, be measured in pertinent experiments. 
 
In Sec.~\ref{sec:Ringinhom}, we  address the question 
of whether the conductance depends on the parameters of the SOC 
{\em solely} through $\cos \lambda_{\text{AC}}$.  
Specifically, we consider a single-channel scattering through a ring interferometer subject to a 
perpendicular  {\it inhomogeneous}  electric field that is responsible for Rashba SOC 
and is controlled by {\it two} different strength parameters $\beta_1$ and $\beta_2$.  
The basic question posed here is whether the conductance $g$ depends {\it independently} on 
$\beta_1,\beta_2$, or only through the ACP [explicitly, through $\cos \lambda_{\text{AC}}(\beta_1,\beta_2)$]. 
First, we derive the closed expressions both for the ACPF and for the ACP 
[concretely $\cos \lambda_{\text{AC}}(\beta_1,\beta_2)$].  
The scattering problem is then solved analytically, 
and, quite remarkably, it is found that the conductance $g(k,\mathcal{X} ; \beta_1,\beta_2)$  
has precisely the {\it same} functional form $\mathcal{G}_{\bigcirc} \left(\cos \lambda_{\text{AC}}(\beta_1,\beta_2); k,\mathcal{X} \right)$  
as in the previous problem (i.e., interferometer under a homogeneous field leading to Rashba SOC specified by a single parameter $\beta$).  
To substantiate this form of universality, we generalize the discussion and consider a system wherein the SOC depends 
on any number of parameters $\{ \beta_{i} \}$ and arrive again at the {\em same} functional form albeit 
with $\cos \lambda_{\text{AC}}$ that now depends on the set of SOC strength parameters.  
The universal function $\mathcal{G}_{\bigcirc} $ given in Eq.~\eqref{eqn:universal-fn-g} and the expression of 
the conductance \eqref{eqn:expression-g-inhom-gen}, or more generally, the observation that, once the geometry is given, 
the conductance $g$ depends on the SOC parameters only through the universal function of the ACP are the central results of this paper.  
We discuss this remarkable observation within the context of non-Abelian SU(2) gauge invariance and show that, in fact, it even goes beyond 
what  gauge-invariance arguments imply.  Generalization to interferometers of arbitrary shapes is discussed as well.  
 
As pointed out above, in addition to the conductance, the electron spin polarization ${\bf P}$ is 
also an experimentally relevant observable whose relation to the ACPF is interesting. 
First, we briefly recall in Sec.~\ref{sec:NoP}, that in a system with two strictly one-dimensional leads 
(i.e., the source and drain), time-reversal invariance implies the absence of electron spin polarization. 
Therefore, in order to study spin polarization, we analyze, in Sec.~\ref{Square}, a tight-binding model for electron transmission through 
a two-channel interferometer of a square form, subject to a 
perpendicular  {\it non-uniform}  electric field controlled by two Rashba SOC strength parameters $\beta_x$ and $\beta_z$. 
It is shown again that the conductance $g$ depends on the SOC parameters  {\it only through} the ACP $\lambda_{\text{AC}}(\beta_x,\beta_z)$. 
On the other hand, the spin polarization $\mathbf{P}$ (that is now finite in this two-channel system) is found to depend {\it separately} 
on $\beta_x$ and  $\beta_z$.  Further analysis shows that the polarization vector ${\bf P}$ depends separately on $\cos \lambda_{\text{AC}}$ 
and on the traceless part of the ACPF. This novel finding indicates that the whole ACPF (i.e., both the traceful and traceless parts) is accessible to measurement in mesoscopic interferometry. 
Quite remarkably, even within such a simple model, it is found that the transmitted spin polarization ${\bf P}^{\text{T}}$ 
reaches high value, close to 40\%. 

A short list of our main results is given in the summary (Sec.~\ref{summary}).
Manipulating non-integrable phase factors in terms of path-ordered integrals and solving the corresponding scattering problems requires technically involved calculations, which are relegated to the appendices.

\section{Formulation}
\label{Sec2}
\subsection{SU(2) vector potential}
\label{subsec-vector-potential}
The main assumption employed is that the SOC enters
the kinetic energy operator through an SU(2) gauge field\cite{{Anandan-89}}\cite{Frohlich-S-93} ${\bm A}$  
whose precise form depends on the details of the SOC mechanism. 
For example, in vacuum, the Pauli equation implies 
${\bm A} \equiv \tfrac{\hbar}{4 mc}{\boldsymbol \sigma} \times {\bf E}$ where $m$ is the electron mass,  
$\boldsymbol{\sigma}=(\sigma_{x},\sigma_{y},\sigma_{z})$ is the vector of the three Pauli matrices, and ${\bf E}$ is the local electric field. 
The SU(2)-invariant Schr\"odinger equation for a free electron (with charge $-e$ and energy $E$) is given by:
\begin{equation} \label{1}
\left({\bf k}+\tfrac{e}{\hbar c} {\bm A} \right)^2 \psi(\bfr)=\veps \psi(\bfr),
\end{equation}
where ${\bf k}=-i {\bm \nabla}$, $\veps=\frac{2mE}{\hbar^2}$
and $\psi(\bfr)$ is the electron wave function in the form of a 2-component spinor.  
 If the parenthesis in Eq.~\eqref{1} are opened, the SOC term of the Hamiltonian is proportional to 
 $(\mathbf{k} \times {\bf E}) \cdot {\bm \sigma} \propto {\bf B}_{\text{eff}} \cdot {\bm \sigma}$, 
 where $ {\bf B}_{\text{eff}}$ is an effective magnetic field felt by the electron in its rest frame.  
 In solid-state physics (more specifically, within the physics of semiconductors), this form of SOC, 
 proportional to $(k_x \sigma_y-k_y \sigma_x)$, is related to the Rashba mechanism.  
 For two-dimensional semiconductors,  the Dresselhaus SOC mechanism, proportional to $(k_x \sigma_x-k_y \sigma_y)$  corresponds to the case 
 $ {\bf B}_{\text{eff}} \propto (k_x,-k_y,0)$ 
 is also relevant.   
In our examples presented below, 
the explicit expressions for the ACPF are derived for both Rashba and Dresselhaus SOC mechanisms. 
\subsection{SU(2)-invariant Schr\"odinger equation for electron on a ring}
\label{Ring}
To introduce the notions of the ACPF/ACP in mesoscopic physics, let us consider (as an example) an electron moving on a metallic ring 
of radius $R$ lying on the plane $z=0$
centered at the origin of the $x$-$y$ plane. 
Its position is specified by the polar vector 
 ${\bf r}=(R,\theta)$ (see Fig.~\ref{fig:ring}). 
The electron is subject to an electric field ${\bf E} \perp \widehat{\bm \theta}$. 
Introducing $R$ as the length unit enables us to rewrite the Schr\"odinger equation \eqref{1} in terms of dimensionless quantities.
Within the SU(2) 
formulation of the Pauli equation \cite{Anandan-89,Frohlich}, 
the SU(2) vector potential in the ring geometry  takes the form, 
\vspace{-0.in}
\begin{equation} \label{2a}
{\bm A}(\theta)=\beta(\theta) \widehat{\bf E}(\theta)\times {\bm \sigma},
 \vspace{-0.03in}
\end{equation}
where the real parameter $\beta(\theta)$ specifies the strength of the local SOC. 
For example, if the Rashba Hamiltonian is written as 
$H_{\text{R}} =i \alpha_{\text{R}}   {\bm \sigma} {\cdot} ({\bf E} {\times} {\bm \nabla})$   
(where $\alpha_{\text{R}}$ is the Rashba SOC parameter), 
then $\beta(\theta)=\frac{mR \alpha_{\text{R}}}{\hbar^2} \vert {\bf E}(\theta) \vert$. 

In the present ring geometry, ${\bf k}$ is parallel to the tangential unit vector $\widehat{\bm \theta}$ 
and we are concerned only with the tangential component of 
$ {\bm A}(\theta)$, that is an element of the su(2)-algebra, 
\begin{equation} \label{calA} 
 {\mathcal A}(\theta)={\bm A}(\theta)\cdot \widehat{\bm \theta} 
 =\beta(\theta) [ \widehat{\bm \theta} \times \widehat{\mathbf{E}} ]   \cdot {\bm \sigma} 
 \equiv \beta(\theta)\widehat{\bf n}\cdot {\bm \sigma} 
 \;  \left[ \in \mbox{su(2)}  \right].
 \end{equation} 
 Classically, $ \widehat{\bm \theta} \parallel {\bf v}$ (where ${\bf v}$ is the electron velocity)
so that  $\widehat{\bf n}$ points along ${\bf E} \times {\bf v}$ that is the 
effective magnetic field ${\bf B}_{\text{eff}}$
felt by the electron in its rest-frame. Quantum mechanically, $\widehat{\bf n}$ encodes the 
nature of the SOC.  
Specifically, $\widehat{\bf n} = \widehat{\bm \theta} \times \widehat{\mathbf{E}}$ for Rashba SOC, 
while for two-dimensional Dresselhaus SOC, 
$\widehat{\bf n}$ is calculated through  
$\widehat {\bf n} \cdot {\bm \sigma} = - (\widehat{\bm \theta})_x \sigma_x + (\widehat{\bm \theta})_y \sigma_y$ 
With the unit vector $\widehat {\bf n}$, 
the SU(2)-invariant Schr\"odinger equation (in dimensionless units) for  the spinor $\psi(\theta)=\binom{\psi_\ua(\theta)}{\psi_\da(\theta)}$ is written as:
\begin{equation} 
\label{dlse}
\left[-i \frac {d}{d\theta}+ \beta (\theta) \widehat{\bf n}(\theta) \cdot {\bm \sigma} \right]^2 \psi(\theta)=\veps \psi(\theta) \; ,
\end{equation} 
where $\veps=\frac{2 m R^2}{\hbar^2} {\cal E} \equiv k^2$, with ${\cal E}$ being the electron energy. 
 \subsection{Definition of ACPF and ACP}
 \label{subsec-ACPF-ACP}
If an electron residing on this ring  moves adiabatically from $\theta$ to $\theta+ d\theta$  
along an infinitesimal directed arc $d{\bm \ell}=R \hat{\bm \theta} d \theta$, it gains an  SU(2) {\it matrix-valued} phase factor according to
\begin{equation} \label{2}
\psi(\theta+d \theta)= \be^{i \beta (\theta) \widehat{\bf n}(\theta) \cdot {\bm \sigma} d \theta} \psi(\theta)   \; .
\end{equation}    
When, the SOC strength $\beta$ and the unit vector $\hat {\bf n}$ depend on the position $\theta$, two phase factors do {\em not} commute.  
Thus, the SU(2) phase factor accumulated from $\theta=0$ up to a {\it finite angle} $\theta$ is given by 
the following {\it path-ordered} (Wilson) integral (or equivalently, an ordered product of infinitesimal phase factors)\cite{Wu-Yang,Anandan-89}: 
\begin{equation} 
\begin{split}
\psi[\theta] &= \mathcal{P}  \exp \left\{ \int_{0}^{\theta}
 i \beta(\theta^{\prime}) \widehat{\bf n} (\theta^{\prime}) {\cdot} {\bm \sigma} d \theta^{\prime}  \right\} \psi(0)  \\
&= \lim_{N \to \infty} \left\{ \prod_{\substack{j=1\\ \longleftarrow }}^{N}
\be^{ i \beta(j \Delta \theta) \widehat{\bf n}(j \Delta \theta) {\cdot} {\bm \sigma} 
\Delta \theta  } \right \} \psi(0)
\equiv F_{\text{AC}}[\theta ;\beta] \psi(0),
\end{split}
\label{eqn:Wilson-line-discretized}
\end{equation}
where $\Delta \theta =\theta/N$ and the arrow $\longleftarrow$ means that matrices are multiplied from the right to the left. 
Strictly speaking, 
the path-ordered integral and related quantities are all {\it functionals} of the local SOC strength $\beta(\theta)$.  
To save on notation, however, we will simply write $\beta$ as an argument and omit its $\theta$-dependence unless it is necessary.   

Note that, in fact, $F_{\text{AC}}[\theta ;\beta]$ is an SU(2) gauge transformation such that,  
after the substitution $\psi(\theta) \to F_{\text{AC}}[\theta ;\beta] \xi(\theta)$  in the Schr\"odinger equation \eqref{dlse}, the SU(2) vector potential 
$\beta (\theta) \widehat{\bf n}(\theta) \cdot {\bm \sigma}$ is {\it locally} eliminated and 
the function $\xi(\theta)$ satisfies the Schr\"odinger equation without the SU(2) gauge potential. 

Since the functional (or gauge transformation) $F_{\text{AC}}[\theta ; \beta]$ is an SU(2) matrix, it can be written as 
$F_{\text{AC}}[\theta ; \beta] =\be^{i \lambda[\theta, \beta] \widehat{\bolOmega}[\theta, \beta] \cdot {\bm \sigma}}$,  
where $\lambda[\theta, \beta]$ and the unit vector $\widehat{\bolOmega}[\theta,\beta]$ should be calculated 
within a given SOC scheme (see examples below).    
The ACPF is then defined as the SU(2) phase factor acquired along the {\it entire} circle:
\begin{equation}  
\begin{split}
& F_{\text{AC}}[2 \pi; \beta] = \mathcal{P}  \exp \left\{ \oint
 i \beta(\theta^{\prime}) \widehat{\bf n} (\theta^{\prime}) {\cdot} {\bm \sigma} d \theta^{\prime}  \right\}  \\
 & \equiv 
 \be^{i \lambda_{\text{AC}} \widehat{\bf b} \cdot {\bm \sigma}}=
 \underbrace{\cos \lambda_{\text{AC}} \mathbf{1}_{2 \times 2}}_{\text{traceful}}
 +\underbrace{i \sin \lambda_{\text{AC}}
 \widehat{\bf b} \cdot{\bm \sigma}}_{\text{traceless}} \; .
 \end{split}
\label{ACPF}
 \end{equation}
Consequently, three real parameters, i.e., the angle $\lambda_{\text{AC}}$ and the unit vector $\widehat{\bf b}$ determine the ACPF 
unambiguously. Our task is to relate $\lambda_{\text{AC}}$ and $\widehat{\bf b}$ to measurable quantities. 

The procedure of multiplying phase factors as in Eq.~(\ref{eqn:Wilson-line-discretized}) and defining $\lambda_{\text{AC}}$ by \eqref{ACPF} 
can be extended straightforwardly to any continuous closed curve. 
Note that $\lambda_{\text{AC}}$ and $\widehat{\bf b}$ encode the ``history'' of both the local SOC strengths $\beta(\theta)$ 
and the directions of the local effective magnetic field $\widehat {\bf n}(\theta)$ along the curve. 
 To see this, consider the equality defining 
 the intermediate $\lambda$ and $\widehat{\bf n}$ for a product of 
two successive phase factors for SOC strengths $\beta_1, \beta_2$ and local field directions 
$\widehat{\bf n}_1, \widehat{\bf n}_2$. Equivalently, one is asked to determined $\lambda$ 
and $\widehat{\bf n}$ in terms of $\beta_1,\beta_2,\widehat{\bf n}_1, \widehat{\bf n}_2$ 
using the following relation,  
\begin{equation}
\begin{split}
& \be^{i \beta_1\widehat{\bf n}_1 \cdot {\bm \sigma}} \be^{i \beta_2\widehat{\bf n}_2 \cdot {\bm \sigma}}  \\
& = (\cos \beta_1 +i \sin \beta_1  \widehat{\bf n}_1 \cdot {\bm \sigma}) 
(\cos \beta_2 +i \sin \beta_2 \widehat{\bf n}_2 \cdot {\bm \sigma})  \\
& =(\cos \lambda +i \sin \lambda \widehat{\bf n} \cdot  {\bm \sigma}) \; .
\end{split}
\end{equation}
With a little effort, one finds 
\begin{equation}
\begin{split}
&  \cos \lambda=\cos \beta_1 \cos \beta_2-\widehat{\bf n}_1 \cdot \widehat{\bf n}_2
\sin \beta_1 \sin \beta_2 \label{cosL},   \ \ 
\\
& \widehat{\mathbf{n}} \sin  \lambda 
= \widehat{\mathbf{n}}_{1} \sin \beta_1 \cos \beta_2 + \widehat{\mathbf{n}}_2 \cos \beta_1 \sin \beta_2  \\
& \phantom{\widehat{\mathbf{n}} \sin  \lambda = } 
- \widehat{\mathbf{n}}_{1}{\times} \widehat{\mathbf{n}}_{2}  \sin \beta_1 \sin \beta_2.
\end{split}
\end{equation}
Therefore, we see that $\lambda$ and $\widehat{\bf n}$ depend 
on the spin-orbit strengths $\beta_1$ and $\beta_2$   
as well as on the corresponding directions $\widehat{\bf n}_1$, $\widehat{\bf n}_2$ of the local effective magnetic fields.  
In particular, $\widehat{\bf n}$ has a component perpendicular to the plane spanned by $\widehat{\bf n}_1$ and $\widehat{\bf n}_2$. 
\subsection{ACPF and ACP}
\label{Subsec-ACPF-ACP}
\begin{figure}[H]
\begin{center}
\includegraphics[width=\columnwidth,clip]{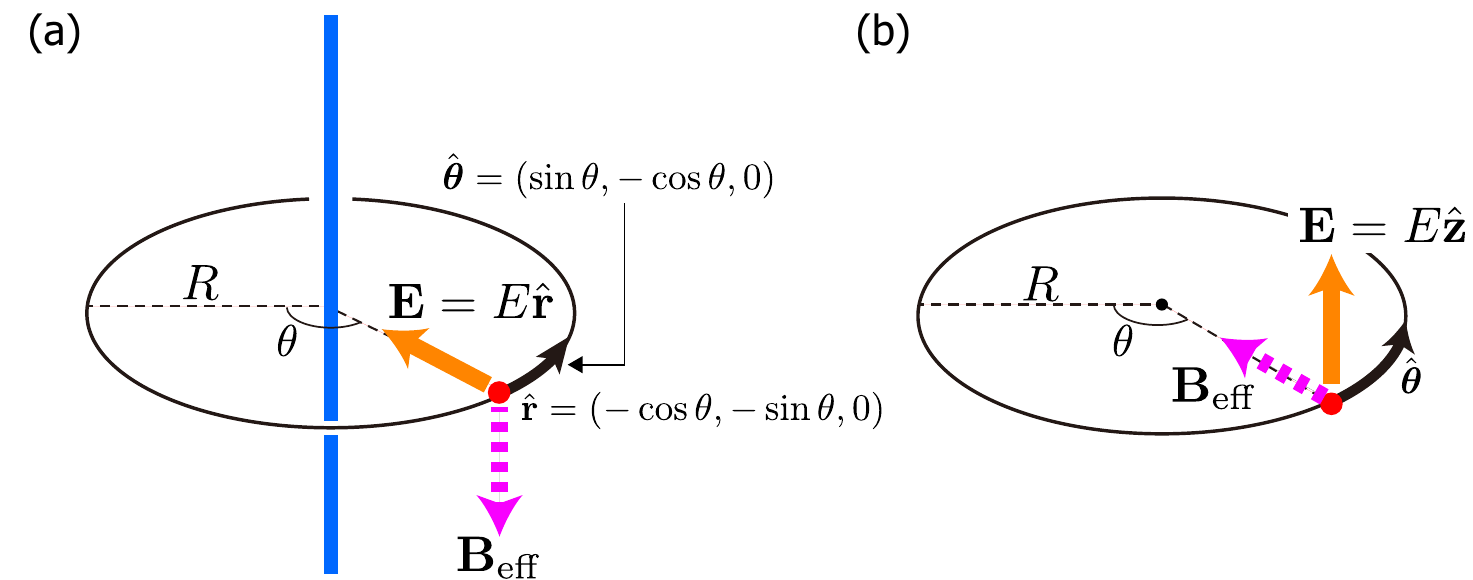}
\end{center}
 \caption{\footnotesize (Color online) 
Two 1D rings (to be used as part of mesoscopic 
interferometer), as considered in the text.  
In (a), a charged particle is subject to a radial electric field $\mathbf{E}=E\hat{\mathbf{r}}$ generated by an infinite uniformly charged wire 
stretched along the ring axis. In this case, ${\bf B}_{\mathrm{eff}}=B_0 \hat{\bf z}$ (independent of $\theta$) 
and a phase factors in Eq.~\eqref{eqn:Wilson-line-discretized} commute. The integral \eqref{ACPF}  
defining the ACPF is then reduced to a simple integral, yielding  
$F_{\text{AC}}[2 \pi; \beta]= \be^{2 \pi i \beta \sigma_z}$ that is $\lambda_{\text{AC}}= 2 \pi \beta$ and 
$\widehat{\bf b}=\hat{\bf z}$. 
In (b), the electric field (generated, e.g, by a gate voltage), is perpendicular to the ring $\mathbf{E}=E\hat{\mathbf{z}}$.   
The electric field generates a position-dependent effective magnetic field $\mathbf{B}_{\text{eff}}$ 
(see Sec.~\ref{sec:Ring}).   
\label{fig:ring}
}
\end{figure}

Formally, being a gauge-invariant quantity, the angle $\lambda_{\text{AC}}$ (or, $\cos \lambda_{\text{AC}}$) 
may be identified with the ACP in our general setting.   This is legitimate since the two eigenvalues of the path-ordered phase factor 
$F_{\text{AC}}[2 \pi; \beta]$ are given by $\be^{\pm i \lambda_{\text{AC}}}$.   
Also, we may call the matrix-valued exponent of the ACPF appearing in Eq.~\eqref{ACPF}  
\begin{equation} 
\varphi_{\text{AC}} \equiv \lambda_{\text{AC}} \widehat{\bf b} \cdot {\bm \sigma} \in 
\mbox{su(2)}
\label{phiAC}
\end{equation} 
the AC generator.   
Traditionally, the ACP is defined as the 
phase acquired by the spin-up component of the spinor moving along the ring 
in the special case of $\widehat {\bf n}$ pointing a fixed direction, e.g., $\widehat {\bf n} =\widehat{\bf z}$, 
that, in this setting, coincides with our $\lambda_{\text{AC}}$.   
Indeed, the familiar (somewhat ubiquitous) example that is used to explain the occurrence of ACP focuses on 
 the special case of a radial electric field $\mathbf{E}=E\widehat{\mathbf{r}}$ generated by an infinite uniformly charged wire \cite{Aharonov-C-84} 
[see Fig.~\ref{fig:ring}(a)], so that the effective magnetic field 
${\bf B}_{\mathrm{eff}}$ is independent of $\theta$
and $\widehat{\bf n} (\theta)=\widehat{\bf z}$.   
Then, all the matrices appearing in the definition \eqref{eqn:Wilson-line-discretized} commute and the path-ordered integral 
reduces to the ordinary integral yielding the following simple results:
\begin{equation}
F_{\text{AC}}[2 \pi; \beta]= \be^{2 \pi i \beta \sigma_z}\, \Rightarrow \  \lambda_{\text{AC}}(\beta) = 2 \pi \beta  \; , \;\; 
\widehat {\bf b} = \widehat{\bf z} \; .
\label{eqn:usual-ACP}
\end{equation}
A beautiful duality between the Aharonov-Bohm and Aharonov-Casher  effects is that the AB phase 
is gained by charged particle circling around a magnetic flux line, while the AC phase is gained by a magnetic moment 
circling around an electric charged line\cite{Rohrlich}. 
As we pointed out above, this duality occurs {\em only in the special case} of constant $\widehat{\bf n}$ 
(e.g., $\widehat{\bf n} =\widehat{\bf z}$) discussed above.  
\subsection{Traceful and Traceless parts of the ACPF}
\label{TraceACPF}
From Eq.~\eqref{ACPF}, we see that
the ACPF $F_{\text{AC}}[2 \pi; \beta]$ contains both the traceful and traceless parts given respectively as, 
\begin{equation} 
\begin{split}
& \cos \lambda_{\text{AC}}= \tfrac{1}{2} \mbox{Tr} \left\{ F_{\text{AC}}[2 \pi; \beta] \right\},   \;   \\
& \sin \lambda_{\text{AC}}\,  \mathbf{b} =-\tfrac{1}{2}i \, \mbox{Tr}\left\{{\bm \sigma} F_{\text{AC}}[2 \pi; \beta] \right\} \; .  
\end{split}
\label{eqn:ACP-vs-F}
\end{equation}
We shall see below that 
the conductance of an electronic (mesoscopic) interferometer is a {\em universal}
function of the traceful part $\cos \lambda_{\text{AC}}$ while the spin polarization vector is determined by {\it both} parts of the ACPF.  
This enables us to measure the full ACPF in experiments.   

\section{Ring Interferometer under perpendicular homogeneous electric field} 
\label{sec:Ring}
In mesoscopic systems, the (dimensionless) conductance $g$ is an excellent tool for 
elucidating the underlying physics, such as the SOC mechanism. Here we consider a ring interferometer wherein electrons are 
subject to strong (Rashba or Dresselhaus) SOC,
 and concentrate on the dependence of the conductance 
on the ACP (more precisely, on $\cos \lambda_{\text{AC}}$). 
This is a classic example discussed by numerous authors where the SOC strength $\beta$ 
is constant (independent of the position) but the SU(2) phase factors $e^{i \beta \widehat{\bf n}(\theta) \cdot {\bm \sigma}}$ 
at different positions $\theta$ now do {\em not} commute since the direction of $ \widehat{\bf n}(\theta)$ varies along the contour. 
One reason for discussing this system here is pedagogical.   First, we explicitly demonstrate how to calculate the path-ordered integrals 
in Eq.~\eqref{eqn:Wilson-line-discretized} when phase factors do not commute and compute the ACP $\lambda_{\text{AC}}$.   
Then, we solve the scattering problem and obtain a simple expression for the conductance 
as a function of the ACP,  thereby substantiate the role of the ACP as a meaningful physical quantity.   

In the problem studied here, the central part of an electronic interferometer is composed of a semiconducting 
one-dimensional ring of radius $R$, placed on the $x$-$y$ plane, and subject to an 
{\em uniform} electric field ${\bf E}= |{\bf E}| \widehat{\bf z}$ [see Fig.~\ref{fig:ring}(b)]. 
Therefore, within the Rashba SOC mechanism, the effective Zeeman field 
$\mathbf{B}_{\text{eff}} \propto \mathbf{E}{\times}\widehat{\bm \theta}$ is now radial and $\theta$-dependent, 
that is, $\widehat{\bf n}(\theta)=\hat {\bfr}=(\cos \theta,\sin \theta,0)$.   
Within the Dresselhaus SOC mechanism, the effective Zeeman field $\mathbf{B}_{\text{eff}}$ 
is also $\theta$-dependent: $\widehat{\bf n}(\theta)= (\sin \theta,\cos \theta,0)$.  
The ACPF pertaining to the ring will be calculated in Sec.~\ref {subsec-ACPF-Ring}. 
In order to measure the conductance through the interferometer, two one-dimensional leads (electrodes), the source and the drain,  
are attached to the ring with an angle $\alpha$ between them (see Fig.~\ref{RingInter}).   
The conductance is calculated by solving the Schr\"{o}dinger equations as will be detailed in Sec.~\ref{sec:scattering-hom}.  
\subsection{The Aharonov-Casher Phase and Phase factor}
\label{subsec-ACPF-Ring}
Before discussing the electron transmission problem,    
let us calculate $F_{\text{AC}}$, which reflects the SOC mechanism along the ring, and  
then extract from it the ACP $\cos \lambda_{\text{AC}}$ upon which the conductance depends.
This is done by directly evaluating the Wilson integral Eq.~\eqref{eqn:Wilson-line-discretized}.   
Other methods for calculating $\lambda_{\text{AC}}$ can be found in,  
e.g.,  Refs.~\citen{Nitta-B-07}, \citen{Wang-V-05}, \citen{Sun-X-W-07}, \citen{Sun-X-W-08}.  
Nevertheless, we are unaware of calculations of the complete matrix-valued ACPF. 

\begin{widetext}
 For Rashba SOC scheme, the analytic form of the 
 phase factor $F_{\text{AC}}[\theta ; \beta(\theta)]$ in Eq.~\eqref{eqn:Wilson-line-discretized}, 
 which is derived in Appendix \ref{app-Ring} [see Eq.~\eqref{eqn:F-theta-beta-formula}], is given for arbitrary angle $0 \le \theta \le 2 \pi$ as:
\begin{equation}
 F_{\text{AC}}(\theta;\beta) \equiv {\cal P} \be^{i  \int_0^{\theta} \beta \widehat{\bf n}(\mu) \cdot {\bm \sigma}  d \mu}    \\
 = 
\begin{pmatrix}
 \frac{ \be^{-\frac{i \theta }{2}} \left\{ 2 y(\beta) \cos \left(y(\beta) \theta \right) +i \sin \left( y(\beta) \theta \right) \right\}}{2y(\beta)} 
& \frac{2 i \, \be^{-\frac{i \theta }{2}} \beta  \sin \left(y(\beta) \theta \right)}{2 y(\beta)} \\
 \frac{2 i \, \be^{\frac{i \theta }{2}} \beta  \sin \left(y(\beta) \theta \right)}{2y(\beta)} 
& \frac{\be^{\frac{i \theta }{2}} \left\{ 2y(\beta) \cos \left( y(\beta) \theta \right) -i \sin \left(y(\beta)\theta \right)\right\}}{2y(\beta)} 
\end{pmatrix},   
\label{eqn:F-theta-beta}
\end{equation}
where $y(\beta) \equiv \frac{1}{2} \sqrt{4\beta^{2}+1}$. 
Then, as noted from Eq.~\eqref{eqn:ACP-vs-F}, $\lambda_{\text{AC}}$ as: 
\begin{equation} 
\cos \lambda_{\text{AC}} =  \tfrac{1}{2}\mbox{Tr}[F_{\text{AC}}(2 \pi; \beta)]  
=  -\cos \left\{ \pi \sqrt{4 \beta^2 + 1} \right\}  = - \cos \left[ 2\pi y(\beta) \right] \; .
\label{eqn:coslambda1}
\end{equation} 
\begin{figure}[h]
\begin{center}
\includegraphics[width=1.2\columnwidth,clip]{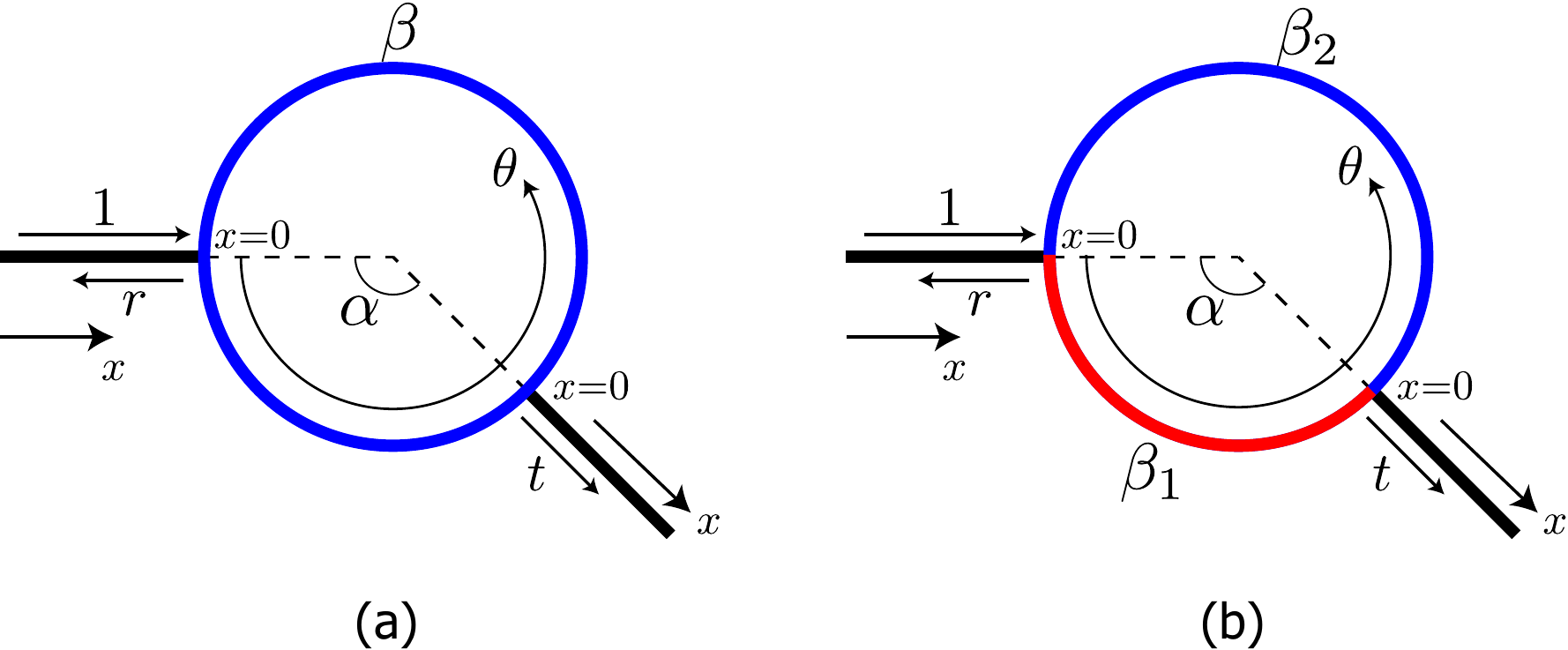}
\end{center}
 \caption{\footnotesize (Color online) 
 Two configurations of 1D ring interferometer (schematic) considered in the text. 
 Electrons approaching the sample from left at polar angle $\theta=0$ are partially reflected 
 and partially transmitted at the second lead at polar angle $\theta=\alpha$ (reflection matrix $r$ 
 and transmission matrix $t$).  The ring is subject to a perpendicular electric field,  
 so that the Rashba SO mechanism generates an effective magnetic field along the radial direction 
 $\hat {\bf n}(\theta)=(\cos \theta, \sin \theta,0)$.  
 (a) The electric field is {\it homogeneous}. The local phase factor for {\it any} polar angle 
 $\theta$ is $\be^{i \beta \, \hat {\bf n}(\theta) \cdot {\bm \sigma}}$. (b)  The electric field is {\it inhomogeneous}.
 The local phase factor for polar angle $0 \le \theta \le \alpha$ is $\be^{i \beta_1 \, \hat {\bf n}(\theta) \cdot {\bm \sigma}}$, 
 while  the local phase factor for polar angle 
 $\alpha \le \theta \le 2 \pi$ is $\be^{i \beta_2 \, \hat {\bf n}(\theta) \cdot {\bm \sigma}}$.  
 Case (a) is thoroughly discussed in the literature and 
displayed in this section for self-contained. Case (b) is novel and discussed in the next section.   %
 \label{RingInter}
}
\end{figure}
\end{widetext}

\begin{figure}[htb]
\centering
\includegraphics[width=0.8\columnwidth,clip]{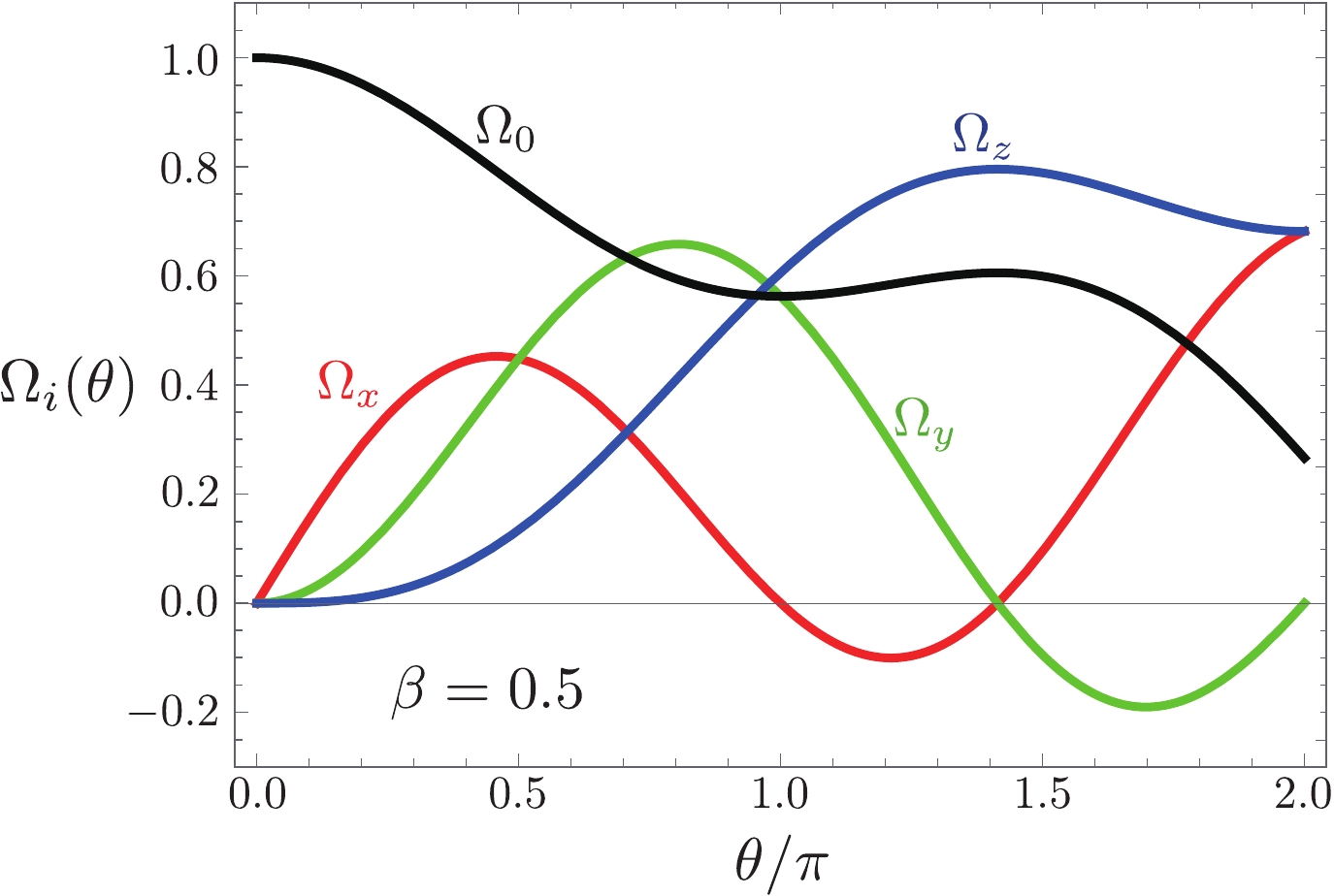}
\caption{(Color online) 
The evolution of the traceful ($\Omega_{0}$) and traceless parts ($\Omega_{x,y,z}$) of $F_{\text{AC}}[\theta;\beta]$ as functions of 
$\theta \in [0,2\pi)$ for the ring under a perpendicular homogeneous field (for $\beta=0.5$). 
Evidently, $\cos \lambda_{\text{AC}} =\Omega_{0}(2\pi)$.   
\label{fig:bxbybz} }
\end{figure}

As has been noted above, the non-Abelian SU(2) phase factor at an arbitrary angel $\theta$
\begin{equation}
F_{\text{AC}}[\theta; \beta]=\cos \lambda(\theta) {\bf 1}_{2 \times 2}
+i \sin \lambda(\theta) \widehat{\bf b}(\theta) \cdot {\bm \sigma}
\end{equation}
consists of both the traceful and traceless parts, and hence the ACP, which is defined by the trace 
of  $F_{\text{AC}}[2 \pi; \beta]$, carries only a part of the full information.  
In this context, it is illuminating to keep track of the evolution of both $\lambda(\theta)$ and $\widehat{\bf b}(\theta)$  
along the contour for $0 \le \theta \le 2 \pi$.  
Specifically, we inspect the followings:
\begin{equation}
\begin{split}
& \Omega_{0}(\theta) \equiv \cos \lambda(\theta)=\tfrac{1}{2} \mbox{Tr} \{ F_{\text{AC}}[\theta; \beta] \}, \\
& \mathbf{\Omega}(\theta) \equiv  \sin \lambda(\theta)\widehat{\bf b}(\theta)=-i \tfrac{1}{2} \mbox{Tr} \{ {\bm \sigma}F_{\text{AC}}[\theta; \beta] \}   \\
& \left[  (\Omega_{0})^{2} + \mathbf{\Omega}^{2} = 1   \right]  \; .
\end{split}
\end{equation}
The $\theta$-dependence of these quantities is displayed in Fig.~\ref{fig:bxbybz}.  

\subsection{The scattering problem and the dimensionless conductance} 
\label{sec:scattering-hom}
Now let us solve the scattering problem to obtain the conductance $g$ for the ring interferometer (radius $R$) 
subject to uniform SOC strength $\beta$ shown in Fig.~\ref{RingInter}(a) (we shall also show later in Sec.~\ref{sec:NoP} 
that the spin polarization $\mathbf{P}$ vanishes for the same setting).  
To write down the Schr\"{o}dinger equation, we use the coordinate $x$ ($\theta$) for the leads (the ring) 
and the radius $R$ as the unit of length [see Fig.~\ref{RingInter}(a) for the definition of the coordinates].   
Accordingly, we use the dimensionless coordinate $x \to x/R$ along either lead and the polar angle $\theta$ along the ring.    
Using the dimensionless wave number $k \to kR$, we can write down the Schr\"{o}dinger equations as: 
\begin{equation} 
\begin{cases} -\frac{d^2 {\bm \Psi}(x)}{d x^2}=k^2 {\bm \Psi}(x),  
\qquad  (\text{$x$ on left or right lead}) \\
\left[  -i \frac{d}{d \theta}-\beta \widehat{\bf n}(\theta) \cdot {\bm \sigma} \right]^2 {\bm \Psi}(\theta)=k^2{\bm \Psi}(\theta)  
\qquad  (\text{$x$ on the ring}) \; .
\end{cases}
\label{eqn:SE}
\end{equation}
To set up the scattering boundary condition, consider an electron with spin projection $\mu= \ua, \da$ approaching the ring from the left lead 
at energy $\veps=k^2$. It is partially transmitted into the right lead with spin projection $\sigma= \ua, \da$ 
and partially reflected back into the left lead with $\sigma= \ua, \da$.   
The corresponding amplitudes $t_{\sigma \mu}$ and $r_{\sigma \mu}$ ($\sigma,\mu= \ua, \da$) are organized into the following $2 \times 2$ 
transmission ($t$) and reflection ($r$) matrices: 
$$r=\binom{r_{\ua \ua} \  r_{\ua \da}}{r_{\da \ua} \ r_{\da \da}} \ ,  \;\;  
t=\binom{t_{\ua \ua} \ t_{\ua \da}}{t_{\da \ua} \ t_{\da \da}}.$$ 
Note the convention that the right (left) spin index corresponds to the initial (final) spin state. 
 The electron wave function at point $\bfr$ (on the ring or on either lead) under the boundary condition that 
 the electron with spin projection $\mu$ impinges on the interferometer from the left is expressed as a spinor: 
$\Psi_\mu(\bfr)=\binom{\psi_{\ua \mu} (\bfr)}{ \psi_{\da \mu} (\bfr) }$.  
Two such spinors for $\mu=\uparrow,\downarrow$ of the incoming electron may be combined 
to form a $2 \times 2$ wave function matrix: 
$$ {\bm \Psi}(\bfr)=\begin{pmatrix} \psi_{\ua \ua} (\bfr) &  \psi_{\ua \da} (\bfr) \\  \psi_{\da \ua} (\bfr) &  \psi_{\da \da} (\bfr) \end{pmatrix}.$$

\begin{widetext}
To  manipulate the formal solutions, it is convenient  to adopt the convention that on both arms, the right-propagating (left-propagating) wave 
is associated with 
$\be^{i k x}$ ($\be^{-i k x}$).  
Then, depending on the region in question, we have the following expressions of the wave function:
\begin{equation} 
{\bm \Psi}(\mathbf{r})=
\begin{cases} 
\be^{i k x} {\bf 1}_{2 \times 2}+ \be^{-i kx} r \quad  (x \in \mbox{left lead}) \\
\be^{i kx} t  \quad  (x \in \mbox{right lead}) \\
F_{\text{AC}}(\theta; \beta)( \be^{i k x} A_+ + \be^{-i kx }A_-) \\
\qquad (0 \le x=2 \pi-\theta \le 2 \pi-\alpha, \ \ \alpha \le \theta \le 2 \pi  \in \text{upper arm of ring}) \\
F_{\text{AC}}(\theta; \beta)( \be^{i k x} B_+ + \be^{-i k x}B_-) \\
\qquad ( 0 \le x=\theta < \alpha  \in \text{lower arm of ring} )  \; ,
\end{cases}
\label{eqn:wave-fn-ring-hom}
\end{equation}
where the origin $x=0$ for the left (right) lead has been taken at the left (right) junction and 
the $x$-direction is chosen in such a way that $k$ is positive for the incoming/transmitted electron  
[see Fig.~\ref{RingInter}(a)].   
The explicit form of the non-Abelian phase factor $F_{\text{AC}}(\theta;\beta)$ has been given already in Eq.~\eqref{eqn:F-theta-beta}.   
\end{widetext}
To compute the wave function requires the knowledge of the six unknown $2 \times 2$ {\it constant} (i.e., $x$, $\theta$-independent) matrices, $A_+,A_-,B_+,B_-,t,r$ that should be 
determined by the following matching conditions.   
By the continuity of the wave function and the current conservation (the Kirchhoff's law) at the left junction ($\theta=0$), we have 
\begin{equation}
\begin{split}
& {\bf 1}_{2 \times 2} + r=F_{\text{AC}}(2 \pi; \beta)(A_++A_-)=B_++B_-,  \\
& {\bf 1}_{2 \times 2}-r= F_{\text{AC}}(2 \pi; \beta)(A_+ -A_-)+(B_+ -B_-) \; ,
\end{split}
\label{eqn:match_left}
\end{equation}
while the matching at the right junction ($\theta=\alpha$) gives:
\begin{equation} 
\begin{split}
t= & F_{\text{AC}}(\alpha; \beta)[\be^{i k (2 \pi -\alpha)}A_+ + \be^{-i k (2 \pi -\alpha)}A_{-}]  \\  
 = & F_{\text{AC}}(\alpha; \beta)[\be^{i k \alpha}B_+ + \be^{-i k \alpha}B_-],  \\
t= & F_{\text{AC}}(\alpha; \beta)[ \be^{i k (2 \pi -\alpha)}A_+ - \be^{-i k (2 \pi -\alpha)}A_-] \\
& +F_{\text{AC}}(\alpha; \beta)[ \be^{i k \alpha}B_+ - \be^{-i k \alpha}B_-].  
\end{split}
\label{eqn:Matchright}
\end{equation}
The matching conditions \eqref{eqn:match_left} and  \eqref{eqn:Matchright} give a set of six coupled equations for the six unknown $2 \times 2$ matrices 
$A_+,A_-,B_+,B_-,r,t$, of which we are interested only in the reflection and transmission amplitude matrices, 
$r=\binom{r_{\ua \ua} \ r_{\ua \da}}{r_{\da \ua} \ r_{\da \da}}$ and $t=\binom{t_{\ua \ua} \ t_{\ua \da}}{t_{\da \ua} \ t_{\da \da}}$. 
\subsection{Conductance }
\begin{figure}[htb]
\centering
\includegraphics[width=0.6\columnwidth,clip]{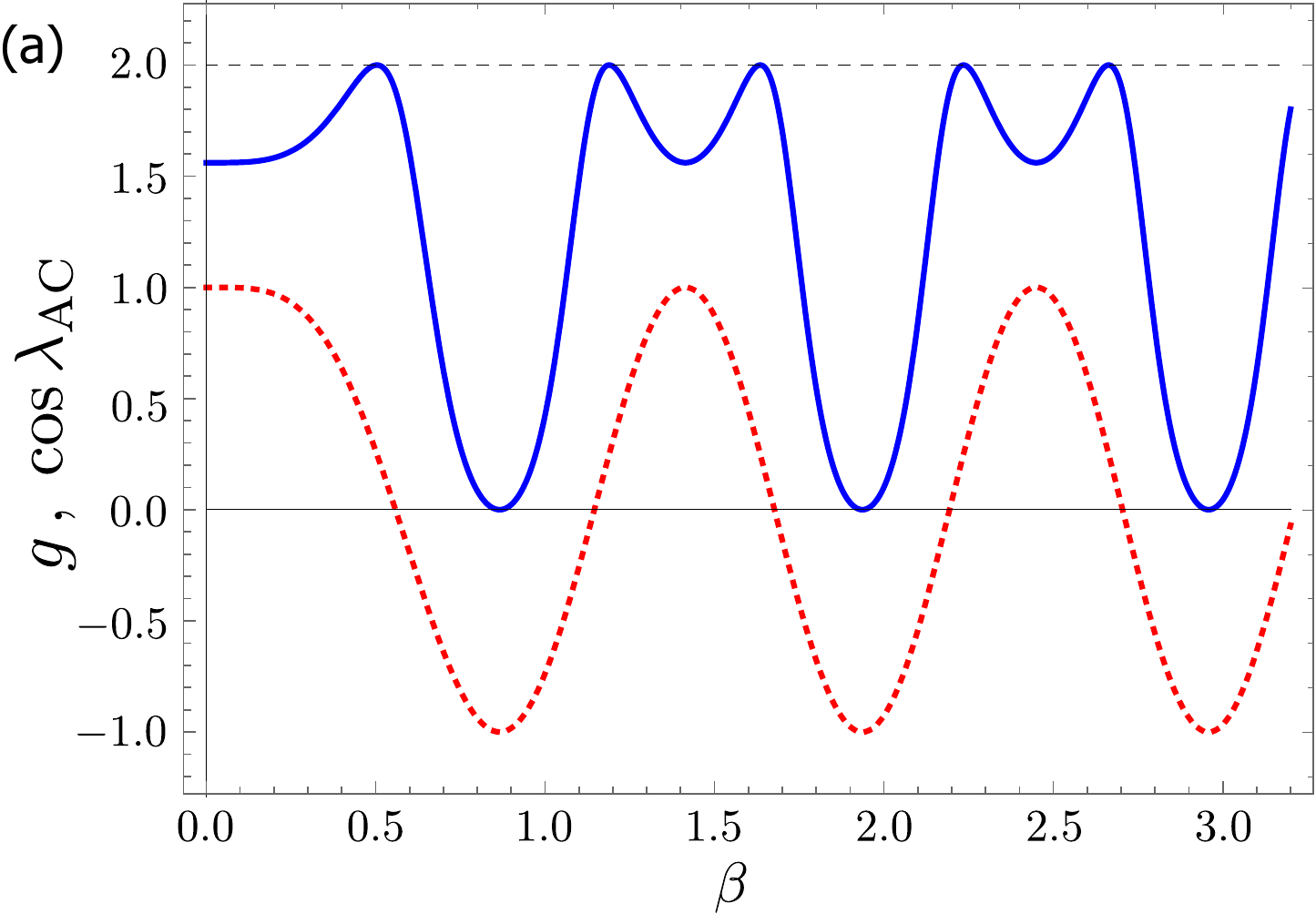}
\hspace{10mm}
\includegraphics[width=0.6\columnwidth,clip]{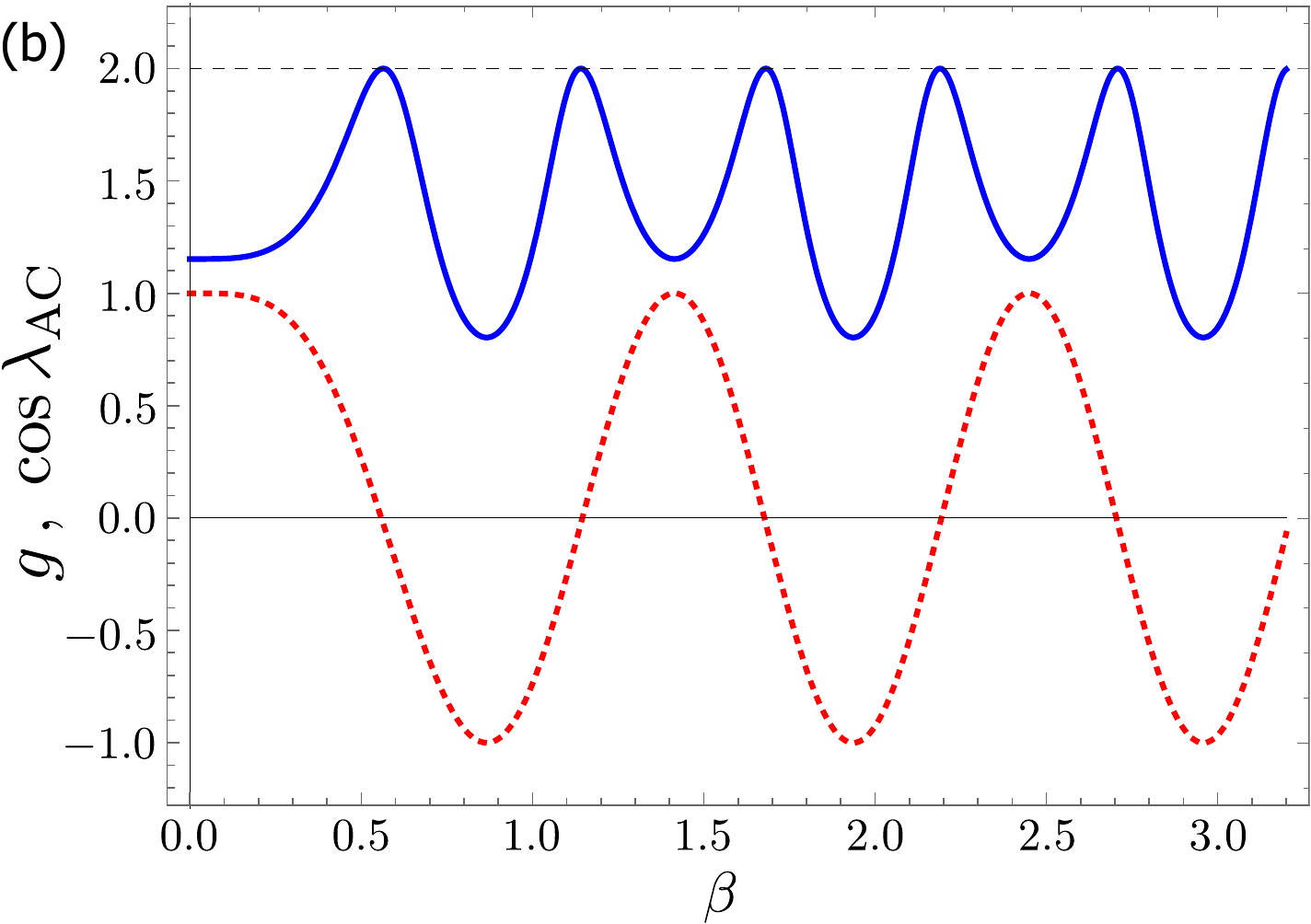} 
\caption{(Color online)  
Plots of the dimensionless conductance $g$ (solid lines) and $\cos \lambda_{\text{AC}}$ (dashed lines) 
as function of spin-orbit strength $\beta$ for the ring interferometer schematically depicted in Fig.~\ref{RingInter}(a).  
In (a), $k=1.25$ and $\alpha=\pi$ (symmetric interferometer) are used, while in (b) $k=1.73$ and $\alpha=1.7$.  
As a property of the closed loop,  $\lambda_{\text{AC}}$ depends neither on $k$ nor on $\alpha$. 
On the other hand, the conductance,
that encodes also the geometric properties of the interferometer 
and the reflection at the junction naturally depends on both $k$ and $\alpha$ [through a universal function \eqref{eqn:universal-fn-g}]. 
}
\label{ACphase}
\end{figure}

For strictly one-dimensional leads,
 the main observable is the dimensionless conductance, $g$ (it will be proved later in Sec.~\ref{sec:NoP} that 
 the spin polarization vanishes identically in this geometry). Due to unitarity and time-reversal symmetry (see below), 
 we actually have the following useful relation:
\begin{equation} 
 g=\mbox{Tr} [t^\dagger t]= 2(1-|r_{\ua \ua}|^2). 
\label{gring1} 
\end{equation} 
\begin{widetext}
As is described in Appendix~\ref{sec:solving-matching-cond}, the above set of equations \eqref{eqn:match_left} and 
\eqref{eqn:Matchright} can be solved for the reflection matrix $r$ as:
\begin{equation}
\begin{split}
r= & \left\{F_{\text{AC}}(2 \pi ;\beta ) + F_{\text{AC}}^{-1}(2 \pi ;\beta ) 
+  \left[ \sin ((2 \pi -\alpha ) k) \sin (\alpha  k)-2 \be^{-2 i \pi  k} \right]  \right\}^{-1} \\
&  \left\{ -F_{\text{AC}}(2 \pi ;\beta ) - F_{\text{AC}}^{-1}(2 \pi ;\beta )+ \left[ \sin ((2 \pi -\alpha ) k) \sin (\alpha  k)+2 \cos (2 \pi  k) \right]   \right\}  \\
= & 
\frac{ - 2 \cos \lambda_{\text{AC}}(\beta) + \left[ \sin ((2 \pi -\alpha ) k) \sin (\alpha  k)+2 \cos (2 \pi  k) \right] }{%
 2 \cos \lambda_{\text{AC}}(\beta)+  \left[ \sin ((2 \pi -\alpha ) k) \sin (\alpha  k)-2 \be^{-2 i \pi  k} \right] }  \mathbf{1}_{2{\times}2}
 \; , 
\end{split}
\label{eqn:r-matrix-ring}
\end{equation}
where we have used Eq.~\eqref{eqn:identity-F-Finv}.   
Combining Eqs.~\eqref{gring1} and \eqref{eqn:r-matrix-ring}, we find a closed expression for $g$: 
\begin{equation}
g (k,\alpha;\beta) =
\frac{16 \sin [(2 \pi -\alpha ) k] \sin (\alpha  k) \left\{ \Phi(\beta) -\cos (2 \pi  k) \right\} + 8 \sin ^2(2 \pi  k)}{
\left\{ 2 \Phi (\beta) +\sin (\alpha k) \sin [(2 \pi -\alpha) k] - 2 \cos (2 \pi  k)\right\}^2+4 \sin ^2(2 \pi  k)}   
\label{eqn:gring}
\end{equation}
where $\Phi (\beta)=\cos \lambda_{\text{AC}}(\beta)=-\cos \pi \sqrt{1+4 \beta^2}$ is derived in Eq.~\eqref{eqn:coslambda1} 
as the traceful part of the non-Abelian ACPF $F_{\text{AC}}(2 \pi; \beta)$.  
For a symmetric interferometer ($\alpha=\pi$), $g (k,\alpha=\pi;\beta)$ reproduces the result of Refs.~\citen{Molnar-P-V-04,Wang-V-05}.  
Thus, we have verified that $g$ is a simple rational function of $\cos \lambda_{\text{AC}}$.   
In fact, as we shall see below, 
the functional form \eqref{eqn:gring} does {\em not} change even if we consider spatially inhomogeneous SOC strengths.  
The conductance $g$ and $\cos \lambda_{\text{AC}}$ for the ring interferometer are displayed in Fig.~\ref{ACphase}(a) and (b) 
for different sets of parameters $(k,\alpha)$.  
\end{widetext}

Equation \eqref{eqn:gring} might be useful for designing the ring interferometer. If the strength $\beta$ of the SOC 
is controllable, then, for a given Fermi energy $k^2$ and interferometer geometry $\alpha$, one can 
tune $\beta$ to obtain the optimal ACP $\lambda_{\text{AC}}(\beta^{\ast})$ at which we achieve 
the maximal conductance $g=2$:
\begin{equation}
\begin{split}
& \cos \lambda_{\text{AC}} (\beta^{\ast}) = -\cos \left(\pi  \sqrt{4 {\beta^{\ast}}^2+1}\right)   \\
& = \frac{1}{2} \left\{ \sin [(2 \pi -\alpha ) k] \sin (\alpha  k)+2 \cos (2 \pi  k) \right\}  \; .
\end{split}
\end{equation}

In Ref.~\citen{Nitta-B-07}, another expression for the conductance is suggested in Eq.~(10) therein, 
which for $\alpha$=$\pi$ reads: $g$=$1$+$\cos \lambda_{\text{AC}}$ 
(the scattering energy is not specified there). We believe that the difference between this expression and the result shown  
in Figs.~\ref{ACphase}(a) and (b) is due to the assumption made in Ref.~\citen{Nitta-B-07} that the contacts 
are adiabatic and no reflection occurs there.  

\section{Ring Interferometer under an inhomogeneous electric field} 
\label{sec:Ringinhom}
In order to partially corroborate our claim that the conductance depends on the 
SOC mechanism {\em only through} the ACP, we now focus on a 
situation where the SOC depends on more than a single strength parameter. Specifically, we first consider the case where there are
two SOC parameters $\beta_1$ and $\beta_2$ that control the conductance through the ACP without touching the geometry or other 
factors (such as energy and geometry, referred to collectively as $X$ in the Introduction).  
 Then, it is required to check whether any two sets of SOC couplings $(\beta_1,\beta_2) \ne (\beta'_1,\beta'_2)$ that give the same 
 ACP [i.e., $\lambda_{\text{AC}}(\beta_1,\beta_2) =\lambda_{\text{AC}}(\beta'_1,\beta'_2)$] yield the identical conductances, 
 that is, $g(\beta_1,\beta_2;X)=g(\beta'_1,\beta'_2;X)$. Returning to the ring interferometer, 
 we consider, as in Fig.~\ref{RingInter}(b), a Gedanken experiment where the SOC strength parameter 
 $\beta_1$ acting on the lower arm  $(0 \le \theta \le \alpha)$ is different from 
$\beta_2$ acting on the upper arm $(\alpha \le \theta \le 2 \pi)$. 
The Schr\"odinger equations \eqref{eqn:SE} for the homogeneous coupling are now modified to incorporate the inhomogeneous SOC coupling as:
\begin{equation} \label{SE1}
\begin{cases} -\frac{d^2 {\bm \Psi}(x)}{d x^2}=k^2 {\bm \Psi}(x)  \quad  ( x \in \, \text{left or right lead})  \; , \\
\left [-i \frac{d}{d \theta}-\beta_1 \widehat{\bf n}(\theta) \cdot {\bm \sigma} \right ]^2 {\bm \Psi}(\theta)=k^2{\bm \Psi}(\theta) \quad 
(\theta \in [0, \alpha]) \; , \\
\left [-i \frac{d}{d \theta}-\beta_2 \widehat{\bf n}(\theta) \cdot {\bm \sigma} \right ]^2 {\bm \Psi}(\theta)=k^2{\bm \Psi}(\theta) \quad 
(\theta \in [\alpha, 2 \pi])  \; . 
\end{cases}
\end{equation}
\ \\
\vspace{-0.4in}
\subsection{Aharonov-Casher  phase factor for a ring under an inhomogeneous electric field}
Let $F_{\text{AC}}(\theta; \beta)$ denote the phase factor for the case of homogeneous field defined in Eq.~\eqref{eqn:F-theta-beta}. 
In Appendix \ref{sec:F-for-two-beta}, we show that the SU(2) phase factor for the case of an {\em inhomogeneous} electric field, 
by which the wave functions ${\bm \Psi}(\theta)$ defined in Eq.~\eqref{SE1} acquire phases along the ring, is given by: 
\begin{equation} \label{Fb1b2}
F_{\text{AC}}(\theta,\alpha; \beta_1,\beta_2)=\begin{cases} F_{\text{AC}}(\theta; \beta_1) \quad ( 0 \le \theta \le \alpha) \; , \\
 F_{\text{AC}}(\theta; \beta_2) \left[F_{\text{AC}}(\alpha; \beta_2)\right]^{-1}F_{\text{AC}}(\alpha; \beta_1)  \\
 \qquad  (\alpha \le \theta \le 2 \pi) \; .
 \end{cases}
 \end{equation}
 It is easy to check that, when $\beta_1 = \beta_2=\beta$, $F_{\text{AC}}(\theta,\alpha; \beta,\beta)=F_{\text{AC}}(\theta; \beta)$ 
 is independent of $\alpha$.  
 The corresponding expression for the ACP is calculated in Appendix \ref{sec:F-for-two-beta} [see Eq.~\eqref{eqn:AC-phase-two-beta}]:
\begin{equation} 
\begin{split}
& \cos [\lambda_{\text{AC}}(\alpha; \beta_1,\beta_2)] =\frac{1}{2}\mbox{Tr}\left [F_{\text{AC}}(2 \pi,\alpha; \beta_1,\beta_2)  \right ] \\
& = \frac{1+4\beta_1 \beta_2}{4 y(\beta_1) y(\beta_2)} 
\sin \left[ \alpha y(\beta_1) \right] \sin \left[ (2\pi-\alpha) y(\beta_2) \right]  \\
& \phantom{=} \;\; 
- \cos \left[ \alpha y(\beta_1) \right] \cos \left[ (2\pi-\alpha)\pi y(\beta_2) \right]   \; ,
\end{split}
\label{lambdab1b2}
\end{equation} 
which, when $\beta_1=\beta_2$, reduces to the ACP for homogeneous field, Eq.~\eqref{eqn:coslambda1}. 

\subsection{The Scattering problem and the conductance}
Now let us calculate the conductance $g$ as a function of $(\beta_1,\beta_2)$.  
As in Sec.~\ref{sec:scattering-hom}, we start by writing down the matching conditions at the two junctions: 
\begin{subequations}
\begin{align}
\begin{split}
& {\bf 1}_{2 \times 2} + r =F_{\text{AC}}(2 \pi; \beta_2)(A_{+}+ A_{-})=B_{+} +B_{-},  \\
& {\bf 1}_{2 \times 2}-r = F_{\text{AC}}(2 \pi; \beta_2)(A_{+} - A_{-})+(B_{+} -B_{-}) \; ,
\end{split}
\label{eqn:match_left-inhom}
\\
\begin{split}
& t= F_{\text{AC}}(\alpha;\beta_2)[ \be^{i k (2 \pi -\alpha)}A_{+} + \be^{-i k (2 \pi -\alpha)}A_{-}]  \\  
& \phantom{t}
=F_{\text{AC}}(\alpha; \beta_1)[\be^{i k \alpha}B_+  + \be^{-i k \alpha}B_-],  \\
& t=F_{\text{AC}}(\alpha;\beta_2)[ \be^{i k (2 \pi -\alpha)}A_{+} - \be^{-i k (2 \pi -\alpha)}A_{-} ] \\
& \phantom{t=}
+F_{\text{AC}}(\alpha;\beta_1)[\be^{i k \alpha}B_+- \be^{-i k \alpha}B_-].  
\end{split}
\label{eqn:Matchright-inhom}
\end{align}
\end{subequations}
\begin{widetext}
If we introduce $\tilde{t}  \equiv F_{\text{AC}}(\alpha; \beta_{1})^{-1} t $ and 
$\widetilde{A}_{\pm} \equiv F_{\text{AC}}(\alpha; \beta_{1})^{-1} F_{\text{AC}}(\alpha; \beta_{2}) A_{\pm}$, 
we can recast \eqref{eqn:match_left-inhom} and \eqref{eqn:Matchright-inhom} into:
\begin{subequations}
\begin{align} 
\begin{split}
\mathbf{1}_{2 \times 2} + r &= F_{\text{AC}}(2 \pi; \beta_{2})F_{\text{AC}}(\alpha; \beta_{2})^{-1} F_{\text{AC}}(\alpha; \beta_{1}) 
( \widetilde{A}_{+} + \widetilde{A}_{-})  \\
&=B_{+} +B_{-} ,  \\
\mathbf{1}_{2 \times 2}-r 
&=  F_{\text{AC}}(2 \pi; \beta_{2})F_{\text{AC}}(\alpha; \beta_{2})^{-1} F_{\text{AC}}(\alpha; \beta_{1}) 
( \widetilde{A}_{+} - \widetilde{A}_{-})+(B_{+} -B_{-})  
\end{split}
\label{eqn:match_left-inhom-2}
 \\
 \begin{split}
\tilde{t} & 
=  \be^{i k (2 \pi -\alpha)}\widetilde{A}_{+} + \be^{-i k (2 \pi -\alpha)}\widetilde{A}_{-}   \\
& =  \be^{i k \alpha}B_{+} + \be^{-i k \alpha}B_{-}  \, ,  \\
\tilde{t} &= 
 \left\{ \be^{i k (2 \pi -\alpha)}\widetilde{A}_{+} - \be^{-i k (2 \pi -\alpha)}\widetilde{A}_{-}  \right\} 
 +  \left\{ \be^{i k \alpha}B_+-\be^{-i k \alpha}B_{-}  \right\} 
 \; .
\end{split}
\label{eqn:Matchright-inhom-2}
\end{align}
\end{subequations}
The matrix $F_{\text{AC}}(2 \pi; \beta_{2})F_{\text{AC}}(\alpha; \beta_{2})^{-1} F_{\text{AC}}(\alpha; \beta_{1}) $ 
appearing in the above equations is nothing but the non-Abelian ACPF $F_{\text{AC}}(2\pi;\alpha;\beta_1,\beta_2)$ 
defined in \eqref{eqn:def-non-Abelian-AB-inhom}.  
Noting that the above set of equations may be obtained by making the replacement: 
\begin{equation}
F_{\text{AC}}^{-1}(\alpha;\beta) t \to  \tilde{t}  \equiv F_{\text{AC}}(\alpha; \beta_{1})^{-1} t  \; , \;\; 
F_{\text{AC}}(2\pi;\beta)  \to F_{\text{AC}}(2\pi;\alpha;\beta_1,\beta_2)
\end{equation}
in those for the homogeneous $\beta$ [i.e., Eqs.~\eqref{eqn:match_left} and \eqref{eqn:Matchright}], 
we can immediately write down the reflection matrix $r$ as:
\begin{equation}
\begin{split}
r =  & \left\{ F_{\text{AC}}(2\pi;\alpha;\beta_1,\beta_2) + F^{-1}_{\text{AC}}(2\pi;\alpha;\beta_1,\beta_2) 
+  \left[ \sin ((2 \pi -\alpha ) k) \sin (\alpha  k)-2 \be^{-2 i \pi  k} \right]  \right\}^{-1} \\
&  \quad \left\{ -F_{\text{AC}}(2\pi;\alpha;\beta_1,\beta_2) - F^{-1}_{\text{AC}}(2\pi;\alpha;\beta_1,\beta_2) 
+ \left[ \sin ((2 \pi -\alpha ) k) \sin (\alpha  k)+2 \cos (2 \pi  k) \right]   \right\}  \\
= & 
\frac{ - 2 \cos[\lambda_{\text{AC}}(\alpha;\beta_1,\beta_2)]  + \left\{ \sin [(2 \pi -\alpha ) k] \sin (\alpha  k)+2 \cos (2 \pi  k) \right\} }{%
 2 \cos[\lambda_{\text{AC}}(\alpha;\beta_1,\beta_2)]  +  \left\{ \sin [(2 \pi -\alpha ) k] \sin (\alpha  k)-2 \be^{-2 i \pi  k} \right\} }  \mathbf{1}_{2{\times}2}
 \; ,
\end{split}
\label{eqn:r-matrix-inhom}
\end{equation}
where the identity \eqref{eqn:identity-F-Finv-inhom} has been used.  
This is exactly the same as the corresponding expression \eqref{eqn:r-matrix-ring} in the case of uniform $\beta$ except that 
now the ACP $\lambda_{\text{AC}}(\beta)$ is replaced with $\lambda_{\text{AC}}(\alpha;\beta_1,\beta_2)$.  
Therefore, we readily obtain the closed-form expression for the conductance $g$:
\begin{equation}
\begin{split}
g (k,\alpha;\beta_1,\beta_2) =
\frac{16 \sin [(2 \pi -\alpha ) k] \sin (\alpha  k) \left\{ \Phi(\alpha;\beta_1,\beta_2) -\cos (2 \pi  k) \right\} + 8 \sin ^2(2 \pi  k)}{
\left\{ 2 \Phi(\alpha;\beta_1,\beta_2) +\sin (\alpha k) \sin [(2 \pi -\alpha) k] - 2 \cos (2 \pi  k)\right\}^2+4 \sin ^2(2 \pi  k)}  
\end{split}
\label{eqn:expression-g-inhom}
\end{equation}
with $\Phi (\alpha;\beta_1,\beta_2) \equiv \cos [\lambda_{\text{AC}}(\alpha;\beta_1,\beta_2) ]$ 
which is given explicitly in Eq.~\eqref{lambdab1b2}.   
The above result suggests us to introduce the following {\em universal} function 
\begin{equation}
\mathcal{G}_{\bigcirc}(\Phi; k,\alpha) 
\equiv  \frac{16 \sin [(2 \pi -\alpha ) k] \sin (\alpha  k) \left\{ \Phi -\cos (2 \pi  k) \right\} + 8 \sin ^2(2 \pi  k)}{
\left\{ 2 \Phi +\sin (\alpha k) \sin [(2 \pi -\alpha) k] - 2 \cos (2 \pi  k)\right\}^2+4 \sin ^2(2 \pi  k)}  
\label{eqn:universal-fn-g}
\end{equation}
that depends only on the Fermi energy $k^{2}$ and the geometrical property $\alpha$, and express $g$ as 
$g (k,\alpha;\beta_1,\beta_2) = \mathcal{G}_{\bigcirc} (\Phi(\alpha;\beta_1,\beta_2); k,\alpha)$.   
The universal function $\mathcal{G}(\Phi; k,\alpha)$ that determines the dimensionless conductance $g$ of the ring interferometer 
is plotted for $\alpha=\pi$ in Fig.~\ref{fig:g-universal}.   
This (partially) proves one of our central claims that $g$ depends on the SOC strength parameters $\beta(\theta)$ 
only through the ACP $\lambda_{\text{AC}}$ for the ring.  

\begin{figure}[htb]
\begin{center}
\includegraphics[width=0.6\columnwidth,clip]{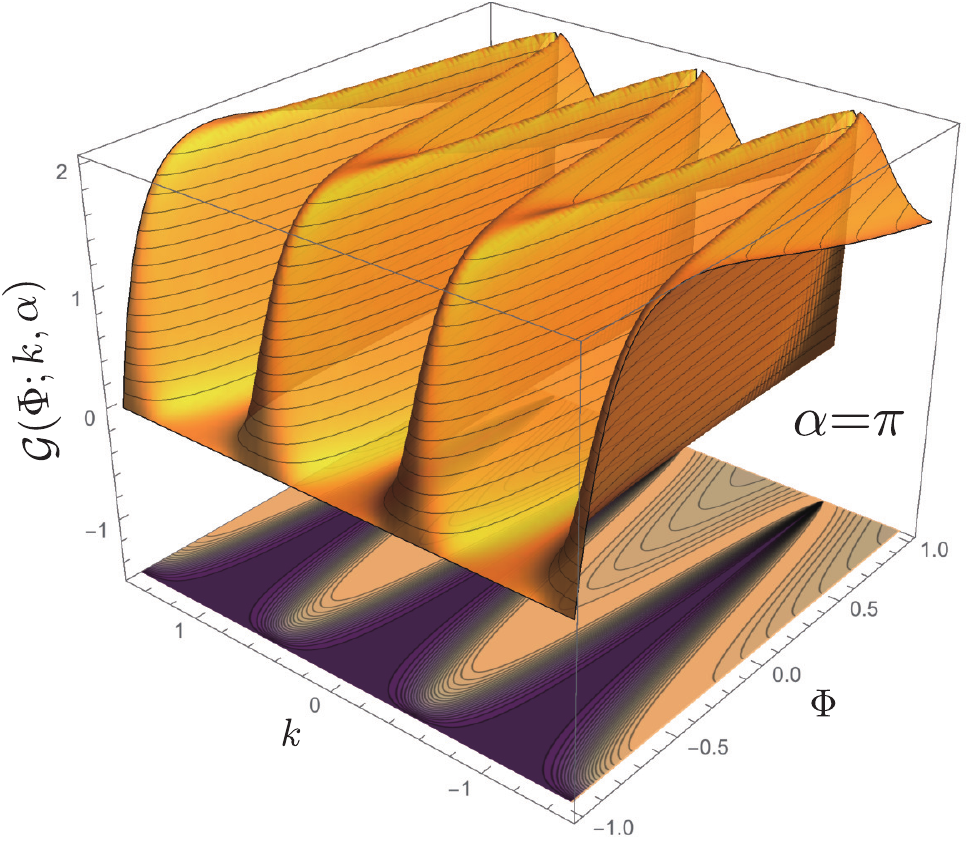}
\end{center}
 \caption{\footnotesize (Color online) 
Plot of the universal function $\mathcal{G}_{\bigcirc} (\Phi; k,\alpha)$ appearing in eq.~\eqref{eqn:universal-fn-g} for $\alpha=\pi$.   %
 \label{fig:g-universal}
}
\end{figure}
\end{widetext}
\subsection{Extension: Multiple intervals with different SOC strengths}
\begin{figure}[htb]
\begin{center}
\includegraphics[width=0.5\columnwidth,clip]{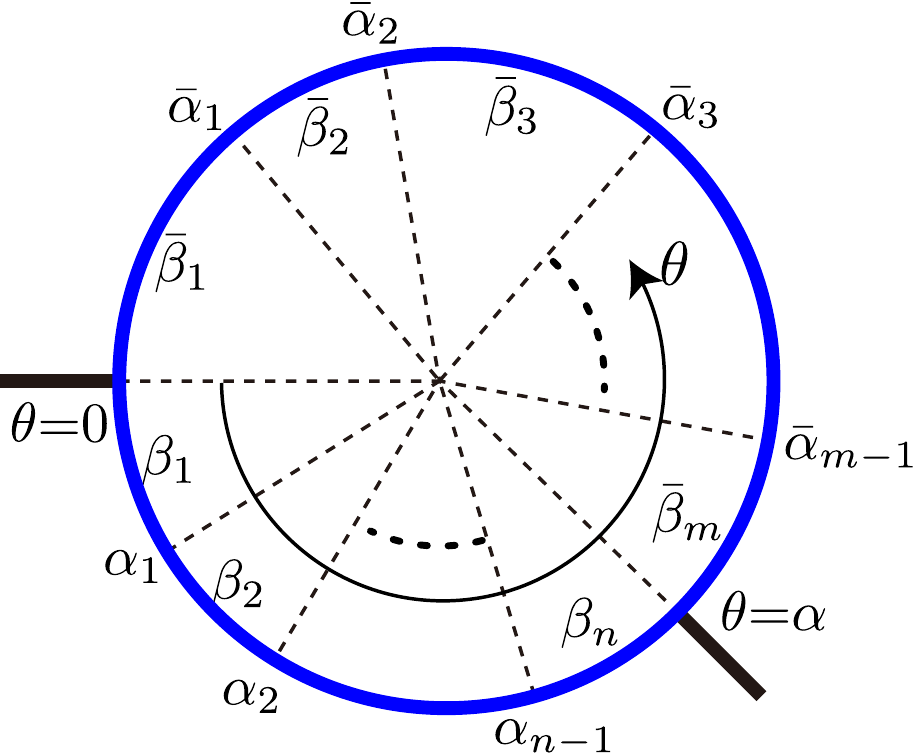}
\end{center}
 \caption{\footnotesize (Color online) 
1D ring interferometer with multiple intervals having different $\{\beta_{i}\}$.   %
 \label{fig:RingInter-gen}
}
\end{figure}

The above observations may be readily generalized to a ring with multiple intervals having an arbitrary number of 
different $\{ \beta_{i}\}$ (see Fig.~\ref{fig:RingInter-gen}).      
To each interval of the ring ($\bar{j}=1,\ldots,m$ for the upper arm, and $j=1,\ldots,n$ for the lower), one of the wave functions of 
the form \eqref{eqn:wave-fn-ring-hom} is assigned with the appropriate value of $\beta_{j}$ (for the lower arm) 
or $\bar{\beta}_{\bar{j}}$ (for the upper arm).  Specifically, we take the following for the lower arm:
\begin{equation}
\begin{split}
& F_{\text{AC}}(\theta;\beta_{j}) \left\{ \be^{ikx} B_{+}^{(j)} + \be^{-ikx} B_{-}^{(j)} \right\}  \\
& (\alpha_{j-1} \leq \theta \leq \alpha_{j}; 
\alpha_{0}=0, \alpha_{n}=\alpha)   
\end{split}
\end{equation}
and similarly for the upper arm.  
The matching conditions at $\theta=\alpha_{i}$ 
($i=1,\ldots,n-1$) or $\theta=\bar{\alpha}_{j}$ ($j=1,\ldots,m-1$) are easily solved to yield the following relations among 
the amplitudes: 
\begin{subequations}
\begin{align}
\begin{split}
& B_{\pm}^{(i+1)} =F_{\text{AC}}^{-1}(\alpha_{i};\beta_{i+1}) F_{\text{AC}}(\alpha_{i};\beta_{i}) B_{\pm}^{(i)}  \\
& \phantom{B_{\pm}^{(i+1)}} 
\equiv G^{-1}(\alpha_{i};\beta_{i},\beta_{i+1}) B_{\pm}^{(i)}  
\end{split}
\label{eqn:wf-adjacent-region-lower}
\\ 
& A_{\pm}^{(j+1)} = G^{-1}(\bar{\alpha}_{j};\bar{\beta}_{j},\bar{\beta}_{j+1}) A_{\pm}^{(j)}  \; .
\label{eqn:wf-adjacent-region-upper}
\end{align}
\end{subequations}
On the other hand, the matching conditions at the two junctions $\theta=0$ and $\theta=\alpha$ are given as:
\begin{subequations}
\begin{align}
\begin{split}
& \mathbf{1}_{2 \times 2} + r =F_{\text{AC}}(2 \pi; \bar{\beta}_{1})(A^{(1)}_{+}+A^{(1)}_{-})  
=B^{(1)}_{+} + B^{(1)}_{-},  \\
& \mathbf{1}_{2 \times 2}-r = F_{\text{AC}}(2 \pi; \bar{\beta}_{1})(A^{(1)}_{+}  - A^{(1)}_{-} )+(B^{(1)}_{+} - B^{(1)}_{-} ) \; ,
\end{split}
\label{eqn:match_left-inhom-2}
\\
\begin{split}
t =& F_{\text{AC}}(\alpha; \bar{\beta}_{m})[\be^{i (2 \pi -\alpha)k} A^{(m)}_{+} +\be^{-i (2 \pi -\alpha)k} A^{(m)}_{-}]   \\
=&  F_{\text{AC}}(\alpha; \beta_{n})[\be^{i k \alpha}B^{(n)}_{+} + \be^{-i k \alpha}B^{(n)}_{-} ],  \\
t =&  F_{\text{AC}}(\alpha; \bar{\beta}_{m})[\be^{i (2 \pi -\alpha)k}A^{(m)}_{+}- \be^{-i (2 \pi -\alpha)k}A^{(m)}_{-}]    \\
& +F_{\text{AC}}(\alpha; \beta_{n})[\be^{i k \alpha}B^{(n)}_{+} - \be^{-i k \alpha}B^{(n)}_{-} ]  \; .  
\end{split}
\label{eqn:Matchright-inhom-2}
\end{align}
\end{subequations}
\begin{widetext}
Now we use the relations \eqref{eqn:F-propagator}, \eqref{eqn:wf-adjacent-region-lower}, and \eqref{eqn:wf-adjacent-region-upper} 
to express $A^{(m)}_{\pm}$ and $B^{(n)}_{\pm}$ in \eqref{eqn:Matchright-inhom-2} by $A^{(1)}_{\pm}$ and $B^{(1)}_{\pm}$: 
\begin{subequations}
\begin{align}
\begin{split}
B_{\pm}^{(n)}  = & G^{-1}(\alpha_{n-1};\beta_{n-1},\beta_{n}) G^{-1}(\alpha_{n-2};\beta_{n-2},\beta_{n-1}) \\
& \cdots   G^{-1}(\alpha_{1};\beta_{1},\beta_{2}) B^{(1)}_{\pm}  \\
\equiv &  F_{\text{AC}}^{-1}(\alpha;\beta_{n}) \left\{ F_{\text{AC}}^{\text{lower}}(\alpha \leftarrow 0) \right\} B^{(1)}_{\pm} 
\end{split}  \label{eqn:Bn-by-B1}
\\
\begin{split}
A_{\pm}^{(m)} & = G^{-1}(\bar{\alpha}_{m-1};\bar{\beta}_{m-1},\bar{\beta}_{m}) 
G^{-1}(\bar{\alpha}_{m-2};\bar{\beta}_{m-2},\bar{\beta}_{m-1}) \cdots G^{-1}(\bar{\alpha}_{1};\bar{\beta}_{1},\bar{\beta}_{2}) A^{(1)}_{\pm}  \\
& \equiv F_{\text{AC}}^{-1}(\alpha;\bar{\beta}_{m}) \left\{ 
F_{\text{AC}}^{\text{upper}}(2\pi \leftarrow \alpha)
\right\}^{-1}
F_{\text{AC}}(2\pi;\bar{\beta}_{1}) A^{(1)}_{\pm} \; ,
\end{split}
\label{eqn:Am-by-A1}
\end{align}
where the SU(2)-valued matrix phases acquired along the lower and upper arms are defined as 
the products of the $\mathcal{F}$ matrices in \eqref{eqn:F-propagator}:
\begin{align}
& F_{\text{AC}}^{\text{lower}}(\alpha \leftarrow 0) \equiv 
\mathcal{F}_{\beta_{n}}(\alpha,\alpha_{n-1})\mathcal{F}_{\beta_{n-1}}(\alpha_{n-1},\alpha_{n-2}) \cdots 
\mathcal{F}_{\beta_{2}}(\alpha_{2},\alpha_{1})\mathcal{F}_{\beta_{1}}(\alpha_{1},0)    \\
& F_{\text{AC}}^{\text{upper}}(2\pi \leftarrow \alpha) \equiv 
\mathcal{F}_{\bar{\beta}_{1}} (2\pi;\bar{\alpha}_{1})   \cdots \mathcal{F}_{\bar{\beta}_{m-1}}(\bar{\alpha}_{m-2},\bar{\alpha}_{m-1})
\mathcal{F}_{\bar{\beta}_{m}}(\bar{\alpha}_{m-1},\alpha)  \; .
\end{align}
\end{subequations}
Plugging \eqref{eqn:Bn-by-B1} and \eqref{eqn:Am-by-A1} into \eqref{eqn:Matchright-inhom-2}, we obtain 
the set of equations identical to \eqref{eqn:match_left-inhom} and \eqref{eqn:Matchright-inhom} 
after the following replacement is made:
\begin{equation}
\begin{split}
& F_{\text{AC}}(2\pi;\beta_{2}) \to F_{\text{AC}}(2\pi;\bar{\beta}_{1})  \\
& F_{\text{AC}}(\alpha;\beta_1)  \to F_{\text{AC}}^{\text{lower}}(\alpha \leftarrow 0)  \; , \;\; 
 F_{\text{AC}}(\alpha;\beta_2)   \to  \left\{ F_{\text{AC}}^{\text{upper}}(2\pi \leftarrow \alpha) \right\}^{-1} F_{\text{AC}}(2\pi;\bar{\beta}_{1}) \; .
\end{split}
\end{equation} 
As the ACPF in Eq.~\eqref{eqn:r-matrix-inhom} is now replaced as:
\begin{equation}
\begin{split}
F_{\text{AC}}(2\pi;\alpha;\beta_1,\beta_2)
&= F_{\text{AC}}(2 \pi; \beta_{2})F_{\text{AC}}(\alpha; \beta_{2})^{-1} F_{\text{AC}}(\alpha; \beta_{1})   \\
& \to 
F_{\text{AC}}^{\text{upper}} (2\pi \leftarrow \alpha) F_{\text{AC}}^{\text{lower}}(\alpha \leftarrow 0) 
\equiv F_{\text{AC}}(\{\alpha_{i}\};\raisebox{-0.3ex}{\rotatebox{90}{$\circlearrowleft$}})  \; ,
\end{split}
\end{equation}
the reflection matrix $r$ is immediately obtained as [see Eq.~\eqref{eqn:r-matrix-inhom}]:
\begin{equation}
\begin{split}
r =  & \left\{ F_{\text{AC}}(\{\alpha_{i}\};\aclockwise) + F^{-1}_{\text{AC}}(\{\alpha_{i}\};\aclockwise) 
+  \left[ \sin ((2 \pi -\alpha ) k) \sin (\alpha  k)-2 \be^{-2 i \pi  k} \right]  \right\}^{-1} \\
&  \quad \left\{ - F_{\text{AC}}(\alpha;\aclockwise) - F^{-1}_{\text{AC}}(\alpha;\aclockwise)
+ \left[ \sin ((2 \pi -\alpha ) k) \sin (\alpha  k)+2 \cos (2 \pi  k) \right]   \right\}   \; .
\end{split}
\label{eqn:r-ring-inhom-general}
\end{equation}
\end{widetext}
Note that $F_{\text{AC}}(\{\alpha_{i}\};\raisebox{-0.3ex}{\rotatebox{90}{$\circlearrowleft$}})$ is the total non-Abelian phase along the ring 
that depends on the positions $\{\alpha_{i}\}$ of the intervals.   
Since $F_{\text{AC}}(\{\alpha_{i}\};\aclockwise)$ is an SU(2) matrix, $F_{\text{AC}}(\alpha;\aclockwise) + F^{-1}_{\text{AC}}(\alpha;\aclockwise)$ 
is a real scalar matrix:
\begin{equation}
F_{\text{AC}}(\{\alpha_{i}\};\aclockwise) + F^{-1}_{\text{AC}}(\{\alpha_{i}\};\aclockwise)  
= \text{Tr}\, F_{\text{AC}}(\{\alpha_{i}\};\aclockwise) \mathbf{1}_{2{\times}2} \; .
\end{equation}
Defining the ACP $\cos[ \lambda_{\text{AC}}(\{\alpha_{i}\};\aclockwise)]$ as before
\begin{equation}
\begin{split}
& \cos[ \lambda_{\text{AC}}(\{\alpha_{i}\};\aclockwise)] \equiv \frac{1}{2}  \text{Tr}\, F_{\text{AC}}(\{\alpha_{i}\};\aclockwise) \\
&= \frac{1}{2}  \text{Tr} \left\{  F_{\text{AC}}^{\text{upper}}(2\pi \leftarrow \alpha) F_{\text{AC}}^{\text{lower}}(\alpha \leftarrow 0) \right\}  
\equiv  \Phi(\{\alpha_{i}\};\aclockwise) \; ,
\end{split}
\end{equation}
we see, from Eq.~\eqref{eqn:r-ring-inhom-general}, that the $r$ is a scalar matrix and depends on $\{\beta_{i}\}$ and $\{\bar{\beta}_{i}\}$ only 
through a single number $\Phi (\{\alpha_{i}\};\aclockwise) = \cos[\lambda_{\text{AC}}(\{\alpha_{i}\};\aclockwise)]$:
\begin{equation}
\begin{split}
&r =  \\
& \left\{ 2 \Phi(\{\alpha_{i}\};\aclockwise) 
+  \left[ \sin ((2 \pi -\alpha ) k) \sin (\alpha  k)-2 \be^{-2 i \pi  k} \right]  \right\}^{-1} \\
&  \left\{ - 2 \Phi (\{\alpha_{i}\};\aclockwise) 
+ \left[ \sin ((2 \pi -\alpha ) k) \sin (\alpha  k)+2 \cos (2 \pi  k) \right]   \right\}  \mathbf{1}_{2{\times}2}  \; ,
\end{split}
\end{equation}
and that so does the conductance $g=\text{Tr}(1-r^{\dagger}r)$.   
This universality of $g$ may be attributed to the fact that, with an appropriate change of variables, we can 
reduce the matching conditions to Eqs.~\eqref{eqn:match_left} and \eqref{eqn:Matchright} for the homogeneous $\beta$ with 
$F_{\text{AC}}(\alpha; \beta)$ replaced with the ACPF $F_{\text{AC}}(\{\alpha_{i}\};\raisebox{-0.3ex}{\rotatebox{90}{$\circlearrowleft$}})$ 
for the given set of $\{\beta_i \}$.  
That $r$ is a scalar matrix is not peculiar to the ring geometry considered in this section.  In fact, as will be shown in Sec.~\ref{sec:NoP}, 
this is generically the case for {\em any} interferometer with two one-dimensional leads.   

Now let us generalize the above observations to cases with arbitrary (presumably continuously varying) distributions of $\beta$.   
If we assume slowly varying $\{\beta_{i}\}$, we may take the continuum limit and introduce 
\begin{equation}
\begin{split}
& F_{\text{AC}}[\beta(\theta);\aclockwise]
= \mathcal{P} \exp \left\{ 
i \oint \!  \beta (\theta^{\prime}) \hat{\mathbf{n}} (\theta^{\prime}) {\cdot} \boldsymbol{\sigma} 
d\theta^{\prime}   \right\}   \\
& \Phi[\beta(\theta);\aclockwise] \equiv \frac{1}{2}  \text{Tr}\, F_{\text{AC}}[\beta(\theta);\aclockwise]    
\end{split}
\label{eqn:F-propagator-path-ordered-inhom}
\end{equation}
for generic distributions $\beta(\theta)$ of the SOC strength [see Eq.~\eqref{eqn:Wilson-line-discretized} 
for the precise definition of the path-ordered product $\mathcal{P}$].   
We can repeat the same argument to obtain the same equations Eqs.~\eqref{eqn:match_left} and \eqref{eqn:Matchright} 
with the ACPF replaced with $F_{\text{AC}}[\beta(\theta);\aclockwise]$ defined above.  
Now it is evident that the conductance $g$ 
is again expressed by the universal function $\mathcal{G}_{\bigcirc}$ in Eq.~\eqref{eqn:universal-fn-g} as: 
\begin{equation}
g [k,\{\alpha\}; \beta(\theta)] =  \mathcal{G}_{\bigcirc} (\Phi[\beta(\theta);\aclockwise]; k,\alpha)  
\label{eqn:expression-g-inhom-gen}
\end{equation}
with $\Phi(\alpha;\beta_1,\beta_2)$ in \eqref{eqn:expression-g-inhom} now replaced with the functional 
$\Phi [\beta(\theta);\aclockwise]$ for {\em any} smooth distributions $\beta(\theta)$.    
This completes the proof for our statement concerning the dependence of $g$ on the SOC strengths.     

One may think that the same result could have been obtained quickly by using gauge invariance of physical observables.  
In fact, the SU(2) gauge invariance requires that the conductance must be a function of the {\em traceful} part of 
$F_{\text{AC}}[\beta(\theta);\aclockwise]$, i.e., the ACP $ \Phi[\beta(\theta);\aclockwise]= \cos[\lambda_{\text{AC}}(\{\alpha_{i}\};\aclockwise)]$.   
To show this, we first note that any polynomial of $F_{\text{AC}}[\beta(\theta);\aclockwise]$ transforms like 
$(F_{\text{AC}}[\beta(\theta);\aclockwise])^{n} 
\to  \mathcal{U}(\theta=0)^{\dagger} (F_{\text{AC}}[\beta(\theta);\aclockwise])^{n} \, \mathcal{U}(\theta=0)$ 
under a local gauge transformation $\mathcal{U}(\theta) \in \text{SU(2)}$.  Therefore, in order to be a gauge-invariant quantity, 
$g$ must be a function of $\{ \text{Tr}F_{\text{AC}}, \text{Tr}(F_{\text{AC}})^{2}, \ldots \}$.  
Note that, in general, $\text{Tr}(F_{\text{AC}})^{n}$ is not a simple function of $\text{Tr}F_{\text{AC}}$.  
However, for SU(2), elementary trigonometry tells us that 
$\text{Tr}(F_{\text{AC}})^{n}/2=\cos[n \lambda_{\text{AC}}(\{\alpha_{i}\};\aclockwise)]$, which, then, can be expressed  
as a polynomial in $\cos[\lambda_{\text{AC}}(\{\alpha_{i}\};\aclockwise)]$.  This completes the proof of the above statement.  
However, this general argument tells nothing about the functional form through which $g$ depends on the traceful part 
$\cos[\lambda_{\text{AC}}(\{\alpha_{i}\};\aclockwise)]$.   In principle, the function may differ for the case of uniform SOC 
considered in Sec.~\ref{sec:Ring} and that of inhomogeneous SOC in this section. 
In fact, what we have found above is much stronger; 
once we fix the geometry of the interferometer (the shape of the ring and the positions of the leads $\alpha$), 
the functional form is universal (except for the $k$-dependence that can be tuned by the initial condition) 
and does {\em not} depend on the detail of the distribution of $\beta(\theta)$.   

Now it is obvious that we can generalize the above argument to the cases of closed curves of {\em arbitrary} shapes.  
Suppose that we parametrizes the curve by an appropriate coordinate in which, after (locally) removing the ``gauge potential'', 
the electron wave function assumes the plane-wave form.    
Then, following the same steps as before, we can reduce the full matching problem to that at the two junctions.  
The reduced problem can be solved in much the same manner as in the ring interferometer to obtain 
the universal form of the conductance $g[\beta] =\mathcal{G}_{\mathcal{X}}(\Phi[\beta])$ ($\Phi[\beta]$: ACP defined by the path-ordered integral 
along the closed curve) with the universal function $\mathcal{G}_{\mathcal{X}}$ 
depending only on the geometry $\mathcal{X}$ [see, e.g., Eq.~\eqref{eqn:expression-g-inhom-gen}].   

\section{Absence of spin polarization in systems with two 1D leads}  
\label{sec:NoP} 
One of our goals in this work is to check whether {\em both} the conductance and the spin polarization  depend on the SOC parameters 
only through the ACPF parameters, that is, $\cos \lambda_{\text{AC}}$ and  $\sin \lambda_{\text{AC}} 
\widehat{\bf b}$ as defined in Eq.~\eqref{ACPF}.    
As far as the conductance is concerned, 
we have already verified, in Secs.~\ref{sec:Ring} and \ref{sec:Ringinhom}, that this is indeed the case 
(at least for generic interferometers with two leads).   
As to the spin polarization, the situation is a bit more involved. 
In strictly one-dimensional systems (without any closed loops),  
time-reversal invariance implies the vanishing spin polarization because, in the absence of closed loops, any SU(2) vector potential 
can be eliminated {\em globally} by a suitable SU(2) gauge transformation.   
Below, we show that the combination of unitarity and time-reversal invariance allows us to generalize 
this statement {\em even to systems with closed loops} provided that there are only two strictly one-dimensional leads (i.e., the source and drain),  
as in the ring interferometer studied in the previous sections.  
Although this property is well known apparently, we could not find a complete substantiation.  
Therefore, we reproduce the argument here so that the paper may be self-contained. 

The idea is to constrain the possible forms of $r$ and $t$ by time-reversal invariance and unitarity.    
Let us denote by $(t_{\text{L}},r_{\text{L}})$ and ($t_{\text{R}},r_{\text{R}}$) the 2$\times$2 transmission ($t$) and reflection ($r$) matrices 
for an electron approaching the interferometer from the left (lead 1) 
and from the right  (lead 2), respectively. Then, the 4$\times$4 $S$-matrix and its unitarity relation read: 
\begin{equation}
\begin{split}
& S=
\begin{pmatrix}
r_{\text{L}} & t_{\text{R}}  \\
t_{\text{L}} & r_{\text{R}} 
\end{pmatrix} ,  \\
& SS^\dagger
=
\begin{pmatrix}
r_{\text{L}} & t_{\text{R}}  \\
t_{\text{L}}  &  r_{\text{R}}
\end{pmatrix}  
\begin{pmatrix}
r_{\text{L}}^{\dagger} &  t_{\text{L}}^{\dagger}  \\
t_{\text{R}}^{\dagger} & r_{\text{R}}^{\dagger}
\end{pmatrix}
= 
\begin{pmatrix}
\mathbf{1}_{2 \times 2} & \mathbf{0}_{2 \times 2} \\
\mathbf{0}_{2 \times 2}  & \mathbf{1}_{2 \times 2}
\end{pmatrix} \; . 
\end{split}
\label{Sunit}
\end{equation} 
For the off-diagonal 2$\times$2 blocks, we have,
\begin{equation} \label{unit12}
\left[ SS^\dagger \right]_{12}= r_{\text{L}} t_{\text{L}}^{\dagger}+t_{\text{R}}r_{\text{R}}^{\dagger}
={\bf 0}_{2 \times 2}   \; .
\end{equation}
Combining Eqs.~\eqref{Sunit} and \eqref{unit12} with time-reversal invariance implies,
\begin{equation} 
\begin{split}
& \text{Tr}[t_{\text{L}}^\dagger t_{\text{L}} + r_{\text{L}}^\dagger r_{\text{L}} ]=2,   \\
& (t_{\text{R}})_{ \sigma \mu}=(-1)^{\sigma-\mu}(t_{\text{L}})^{\ast}_{\bar{\mu } \bar{\sigma}}, \; 
(r_{\text{L}})_{ \sigma  \mu}=(-1)^{\sigma-\mu} (r_{\text{L}})_{ \bar{\mu }  \bar{\sigma}}, \\
& (r_{\text{R}})_{ \sigma  \mu}=(-1)^{\sigma-\mu}(r_{\text{R}})_{ \bar{\mu }  \bar{\sigma}} \; . 
\end{split}
\label{eqn:unitarity1}
\end{equation}
Here $\sigma, \mu=\pm \tfrac{1}{2}$ and $\bar {\sigma}=-\sigma$.  
The second set of equations is proved as follows.  Let $\hat {S}$ denote the scattering {\em operator} each of whose elements  
forms a $4 \times 4$ $S$-matrix.  The initial ket for an electron with spin projection $\mu=\ua,\da$ coming 
from the left (right) is $\vert k , \mu \ra$ ($\vert -k , \mu \ra$).  
The corresponding final bra is $\la k \sigma \vert$ ($\la -k \sigma \vert$) for the transmitted electron and 
$\la -k \sigma \vert$ ($\la k \sigma \vert$) for the reflected electron.
Accordingly, the elements of the $2 \times 2$ transmission matrices $t_{\text{L}}$ and $t_{\text{R}}$ can be written respectively 
as $\la k \sigma\vert \hat {S} \vert k \mu \ra$ and $\la -k \sigma\vert \hat {S} \vert -k \mu \ra$.   
Similarly, the elements of the $2 \times 2$ reflection matrices $r_{\text{L}}$ and $r_{\text{R}}$ can be written 
as $\la -k \sigma\vert \hat {S} \vert k \mu \ra$ and $\la k \sigma\vert \hat {S} \vert -k \mu \ra$, respectively.   
Following the analysis detailed in Ref.~\citen{Goldberger-W-65},  
the time-reversal symmetry of $\hat {S}$ implies the relations specified above.  
Note that the matrix elements of the transmission matrices are related to those {\em on the other side of the sample}, 
while the matrix elements of the reflection matrices are related to those on the {\em same} side.  
These relations imply that spin-flip is absent in the reflection amplitude for each channel, 
i.e., $\left(r_{\text{L/R}}\right)_{ \nu  \bar {\nu}}=0$, 
and also that the diagonal elements of the reflection matrix are equal: $\left(r_{\text{L/R}}\right)_{ \ua  \ua}=\left(r_{\text{L/R}}\right)_{ \da \da}$.   
More generally, these allow the following parametrization:
\begin{equation} \label{constraintstr}
 r=\rho e^{i \theta} {\bf 1}_{2 {\times} 2}, \ \
 t=\begin{pmatrix} \tau e^{i \alpha}&\eta e^{i \beta}\\ \eta e^{i \gamma} & \tau e^{i \delta} \end{pmatrix}, 
 \ \mbox{with} \ \beta-\alpha=\delta-\gamma+\pi \; ,
\end{equation}
where $\rho, \tau, \eta >0$ and, by unitarity, $\rho^2+\eta^2+\tau^2=1$.  Since $r$ is a scalar matrix,  
the reflected spin polarization vanishes because Tr$[r^\dagger \, {\bm \sigma} \, r]=0$.  
As for the transmitted polarization, the constraint $\beta-\alpha=\delta-\gamma+\pi$ 
on the phases of the matrix elements of $t$ implies the vanishing of the transmitted polarization, i.e., Tr$[t^\dagger {\bm \sigma} t] = 0$. 

\section{Two-channel square interferometer: conductance and spin polarization}
\label{Square}
\begin{figure}[htb]
\begin{center} 
{\includegraphics[width=\columnwidth,clip]{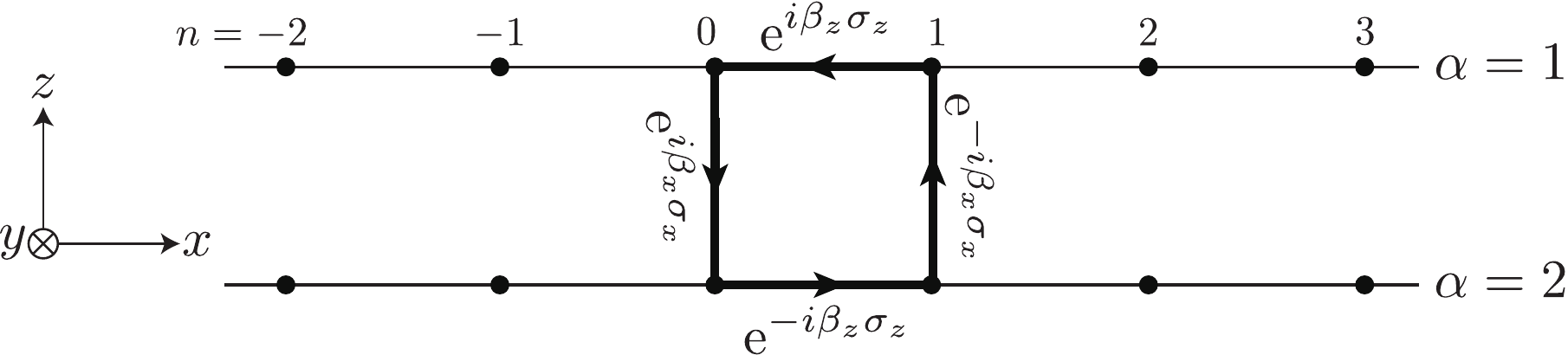}}
\caption{Two-channel tight-binding model for electron scattering from a square square interferometer  lying in the $x$-$z$ plane  (shown by bold line). 
The square part is subject to an {\em inhomogeneous} electric field $E(n,\alpha)\hat {\bf y}$, such that the Rashba SO strength on horizontal and vertical 
links are different. The corresponding SU(2) hopping matrix elements 
are given by $\be^{\pm  i \beta_z \sigma_z}$ 
 along the horizontal links and $\be^{\mp  i \beta_x \sigma_x}$ along the vertical links. 
 Generically, the AC phase , depends   
 on both $\beta_{x}$ and $\beta_{z}$: $\lambda_{\text{AC}}(\beta_x, \beta_z)$.   
However, it is proved analytically and shown graphically below that if $\lambda_{\text{AC}}(\beta_x,\beta_z)= \lambda_{\text{AC}}(\beta'_x,\beta'_z)$ then 
 the corresponding conductances are equal, while the corresponding spin polarizations are not. 
 }
\label{SquareScatt}
\end{center}
\end{figure}

The discussion in the previous section indicates that, in order to analyze the relation 
between the ACPF and spin polarization, 
we need to study an interferometer with more than two 1D leads. 
 In this section, we consider, as a simple example,  
a tight-binding model of electron scattering in a two-channel interferometer 
 having a square geometry displayed in Fig. \ref{SquareScatt}, 
where the local phase factors 
along the links do not commute and the strengths of the SOC along the vertical and horizontal links are not equal.     
The model consists of two chains numbered $\alpha=1,2$ with each chain consisting of integer sites  $-\infty < n < \infty$. The two chains 
are then connected to each other by the two vertical links (rungs) at $n=0$ and $n=1$ (see Fig.~\ref{SquareScatt}).   
The creation and annihilation operators for electrons with spin $\sigma=\uparrow,\downarrow$ are indexed respectively 
as $c^\dagger_{\alpha, n, \sigma}$ and $c_{\alpha, n, \sigma}$, 
and the spin-orbit interaction is active only on the four links forming the square (shown by the bold lines in Fig.~\ref{SquareScatt}).    
The Hamiltonian is written as $H=H_0+H_1$ with 
\begin{equation}
\begin{split}
H_0= & -t \sum_{\alpha=1,2} \sum_{n \le 0}{\bf c}^\dagger_{\alpha, -(n+1)} {\bf c}_{\alpha, -n}  \\
& - t  \sum_{\alpha=1,2}\sum_{n \ge 1}{\bf c}^\dagger_{\alpha, (n+1)} {\bf c}_{\alpha, n}+ \text{h.c.} ,  \\
H_1 =& \sum_{n=0,1}{\bf c}^\dagger_{1,n} \be^{- i \beta_x \sigma_x}{\bf c}_{2,n} 
+\sum_{\alpha=1,2}{\bf c}^\dagger_{\alpha,0} \be^{i \beta_z \sigma_z}{\bf c}_{\alpha,1}+  \text{h.c.} \\
& {\bf c}^\dagger_{\alpha,n} \equiv (c^\dagger_{\alpha,n, \ua},c^\dagger_{\alpha,n, \da}) \; .
\end{split}
\label{H}
\end{equation} 
A similar interferometer with a single channel has been considered 
in Ref.~\citen{Hatano-S-N-07}, where the spin polarization vanishes identically due to the symmetry considerations as has been shown above.  
It is worth mentioning here a few other 
works pertaining to electron spin polarization in mesoscopic interferometers, albeit 
without direct relevance of the ACPF\cite{AA1}\cite{AA2},\cite{AA3}.  

\begin{figure}[htb]
\centering
\centering
{\includegraphics[width=0.45\columnwidth,clip]{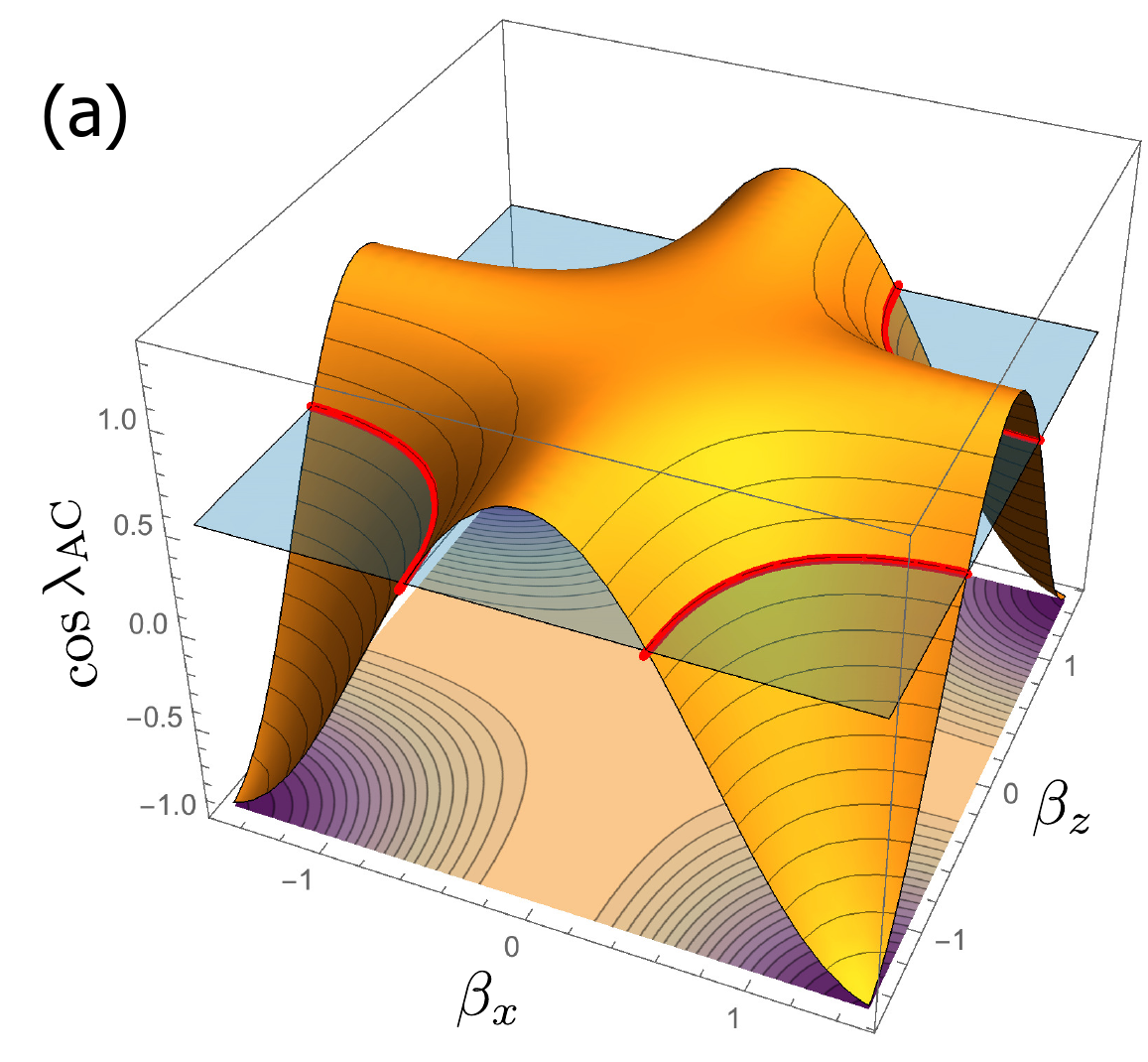}}
\hspace{3mm}
\centering
\includegraphics[width=0.45\columnwidth,clip]{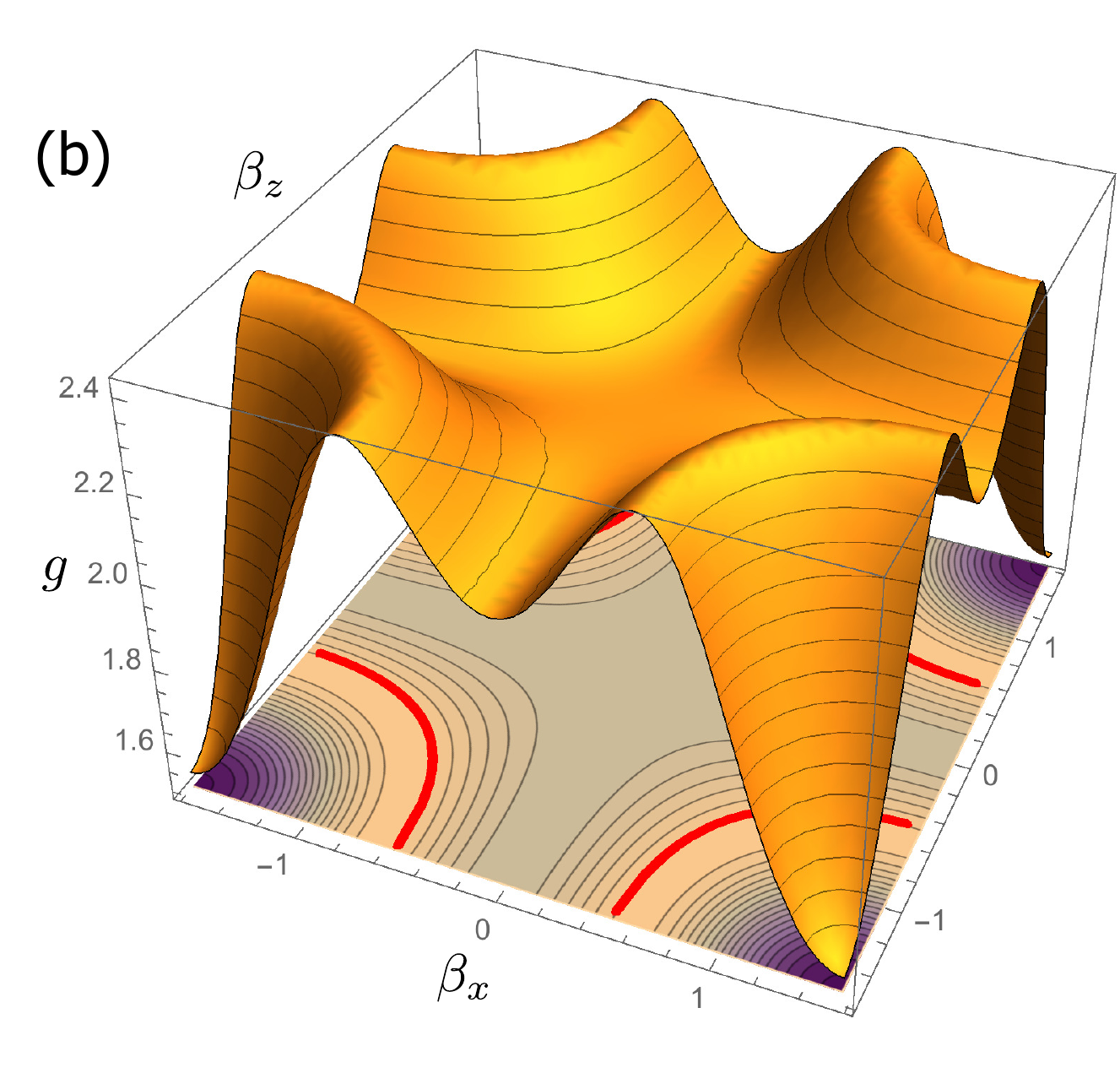}
\caption{(Color online)   
(a) Cosine of ACP $\lambda_{\text{AC}}$ as function of spin-orbit strengths $\beta_x, \beta_z$ 
following Eq.~\eqref{PhaseSquare1}.  Also shown (in blue) is the plane $\cos \lambda=\tfrac{1}{2}$.
For each fixed $\cos \lambda_{\text{AC}}$ there are pairs $(\beta_x,\beta_z)$ on four curves that yield the same 
AC phase $\lambda$. (b) 3D plot of the charge conductance $g(k=0.7,\beta_x,\beta_z)$ [Eq.~\eqref{eqn:gPz-1}] 
of the square interferometer as function of $\beta_x$ and $\beta_z$.  Also plotted are the curves on the $(\beta_x,\beta_z)$ along which $\cos \lambda_{\text{AC}}=1/2=\text{const.}$, namely,  
projection of the red curves in (a) on the $(\beta_x,\beta_z)$ plane.
}
\label{PhaseSquarea2}
\vspace{-0.30in}
\end{figure}

The ACPF  $F_{\mathrm{AC}}$, 
 is a property of the close loop irrespective of the scattering energy $\veps_k=-2 t \cos k$,   
and is given by the product of the four matrices shown in Fig.~\ref{SquareScatt}:
\begin{equation} \label{ACPFSQUARE}
F_{\text{AC}}=  \be^{i \beta_{z}\sigma_z} \be^{- i \beta_{x}\sigma_x}\be^{-i \beta_{z}\sigma_z}\be^{i \beta_{x}\sigma_x}  
\equiv \be^{i \lambda_{\mathrm{AC}}(\beta_x,\beta_z) \widehat{\bf b} {\cdot}\boldsymbol{\sigma}}  \; .
\end{equation}
Consequently, we can express the three parameters by $\beta_x$ and $\beta_z$ as:
\begin{subequations}
\begin{align} 
\begin{split}
& \cos \lambda_{\mathrm{AC}}(\beta_x,\beta_z) =
\tfrac{1}{2}\text{Tr}[F_{\text{AC}}] \\
& = 1-2 \sin^2 \beta_x\sin^2 \beta_z 
\end{split} 
\label{PhaseSquare1} \\
\begin{split}
& \sin  \lambda_{\mathrm{AC}}(\beta_x,\beta_z) \widehat{\bf b}
= -\tfrac{1}{2} i \text{Tr}[{\bm \sigma} F_{\text{AC}}]  \\
& 
=\left (\sin 2 \beta_x \sin^2 \beta_z, \tfrac{1}{2} \sin 2 \beta_x \sin 2 \beta_z, \sin^2 \beta_x \sin 2 \beta_z \right ) \; .
\end{split}
\label{tracelesssquare}
\end{align}
\end{subequations}
The function $\cos \lambda_{\mathrm{AC}}(\beta_x,\beta_z)$ is plotted in Fig.~\ref{PhaseSquarea2}(a), 
together with the plane $\cos \lambda_{\text{AC}}(\beta_x,\beta_z) =\tfrac{1}{2}$.  
According to Eq.~(\ref{PhaseSquare1}), for any fixed value of $\zeta=\cos \lambda_{\text{AC}}$ ($-1 < \zeta < 1$),  
there are four curves $\beta_x=f_i(\beta_z; \zeta)$ [$i=1,2,3,4$; shown by the red curves 
in Fig.~\ref{PhaseSquarea2}(a)] in the square region $[-\tfrac{\pi}{2},  \tfrac{\pi}{2}]{\times}[-\tfrac{\pi}{2},  \tfrac{\pi}{2}]$.  
For all pairs $(\beta_x,\beta_z)$ on these curves, $\cos \lambda_{\text{AC}}$ takes the same value $\zeta$.  

Before presenting the results related to the scattering problem, it is worthwhile 
to consider the eigenvalue problem of the a system of an electron hopping on an isolated square.  
Specifically, we ask if the eigenvalues of the square in Fig.~\ref{SquareScatt}, when it is decoupled from 
the rest of the system, depend on $\lambda_{\text{AC}}$ {\em alone} or separately on $\beta_x$ and $\beta_z$. 
The tight-binding $8 \times 8$ Hamiltonian assumes the following form
\begin{equation} \label{Hsquare}
H_{\square}=\begin{pmatrix} 
0& \be^{i \beta_z \sigma_z}& \be^{-i \beta_x \sigma_x}&0\\
\be^{-i \beta_z \sigma_z}&0&0& \be^{-i \beta_x \sigma_x}\\
\be^{i \beta_x \sigma_x}&0&0& \be^{i \beta_z \sigma_z}\\
0& \be^{i \beta_x \sigma_x}& \be^{-i \beta_z \sigma_z}&0 \end{pmatrix}  \;, 
\end{equation} 
where each entry is a $2 \times 2$ matrix acting on the spinor wave function at each site of the square. Simple calculations 
find the following four different eigenvalues each of which is two-fold (Kramers) degenerate:  
\begin{equation}
\label{eigenvalues-square}
E_\square=\pm 2 \cos \left[ \tfrac{ \lambda_{\text{AC}}(\beta_x,\beta_z)}{4}\right] , \ \ 
\pm 2 \sin \left[ \tfrac{ \lambda_{\mathrm {AC}}(\beta_x,\beta_z)}{4} \right]  \;  .  
\end{equation}
Thus, the eigenvalues depends on $\beta_x$ and $\beta_y$ only through 
$\lambda_{\text{AC}}(\beta_x,\beta_y)$ defined in Eq.~\eqref{PhaseSquare1}. 

The solution of the scattering problem using the 
transfer matrix method is worked out in Ref. \citen{Avishai-B-17}, but for the readers' convenience, 
we detail the solution in Appendix \ref{sec:solution-square}. It yields the $4 \times 4$ (2 for spin $\uparrow/\downarrow$ and 2 for channel $\alpha=1,2$) 
transmission and reflection  matrices $t$ and $r$ whose matrix elements $t_{\alpha' \sigma';\alpha \sigma}$ 
and $r_{\alpha^{\prime} \sigma^{\prime} ;\alpha \sigma}$ 
give the amplitudes of a spin-$\sigma$ electron in channel $\alpha$ transmitted or reflected into a spin-$\sigma^{\prime}$ one 
in channel $\alpha^{\prime}$.     
They depend on the SOC strengths $\beta_x$, $\beta_z$, as well as on 
the wave number $k$ that determines the energy $\veps= - 2 t \cos k$ of the incoming electron.  
 
\begin{figure}[htb]
\centering   
\includegraphics[width=0.45\columnwidth,clip]{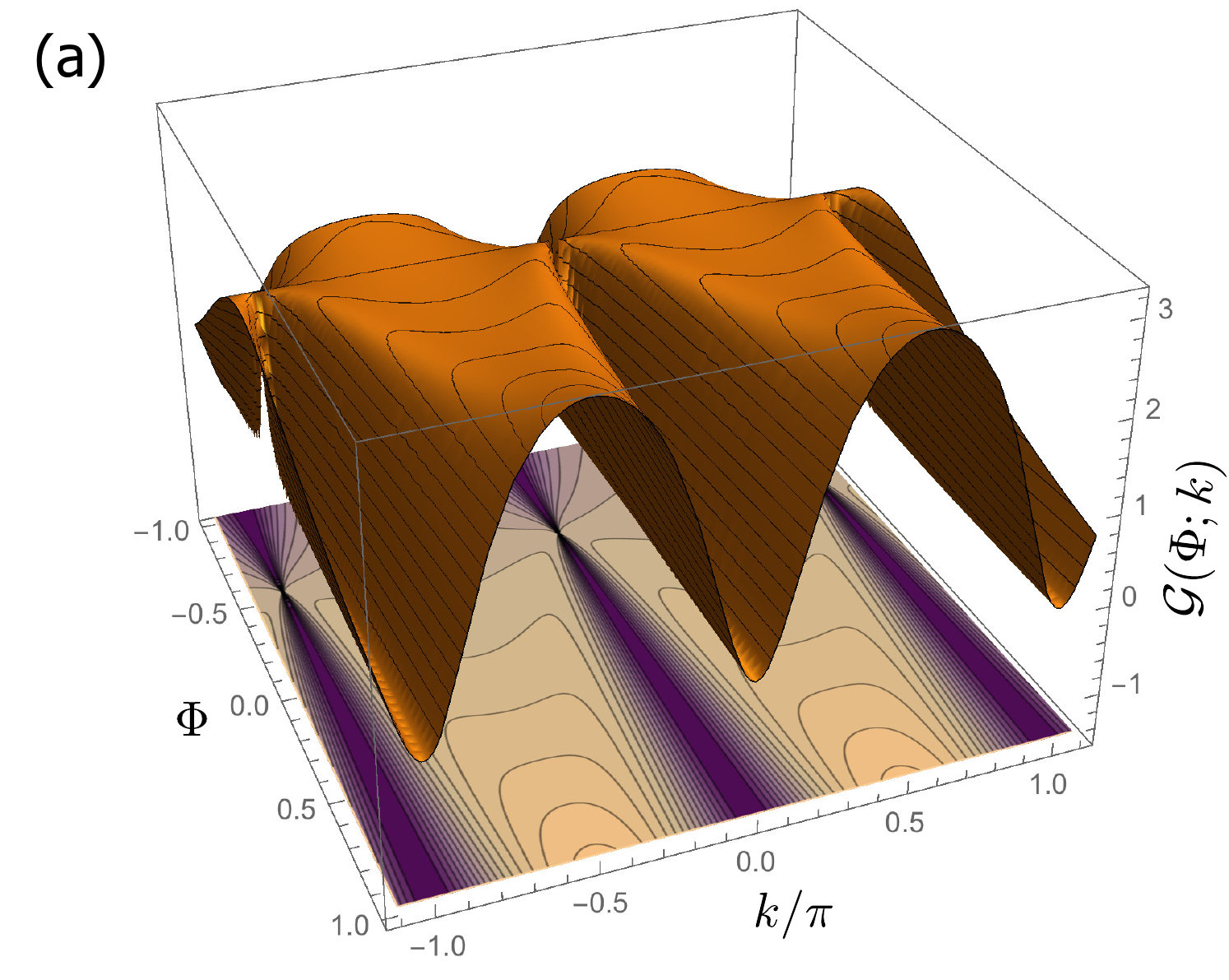}
\hspace{3mm}
\includegraphics[width=0.45\columnwidth,clip]{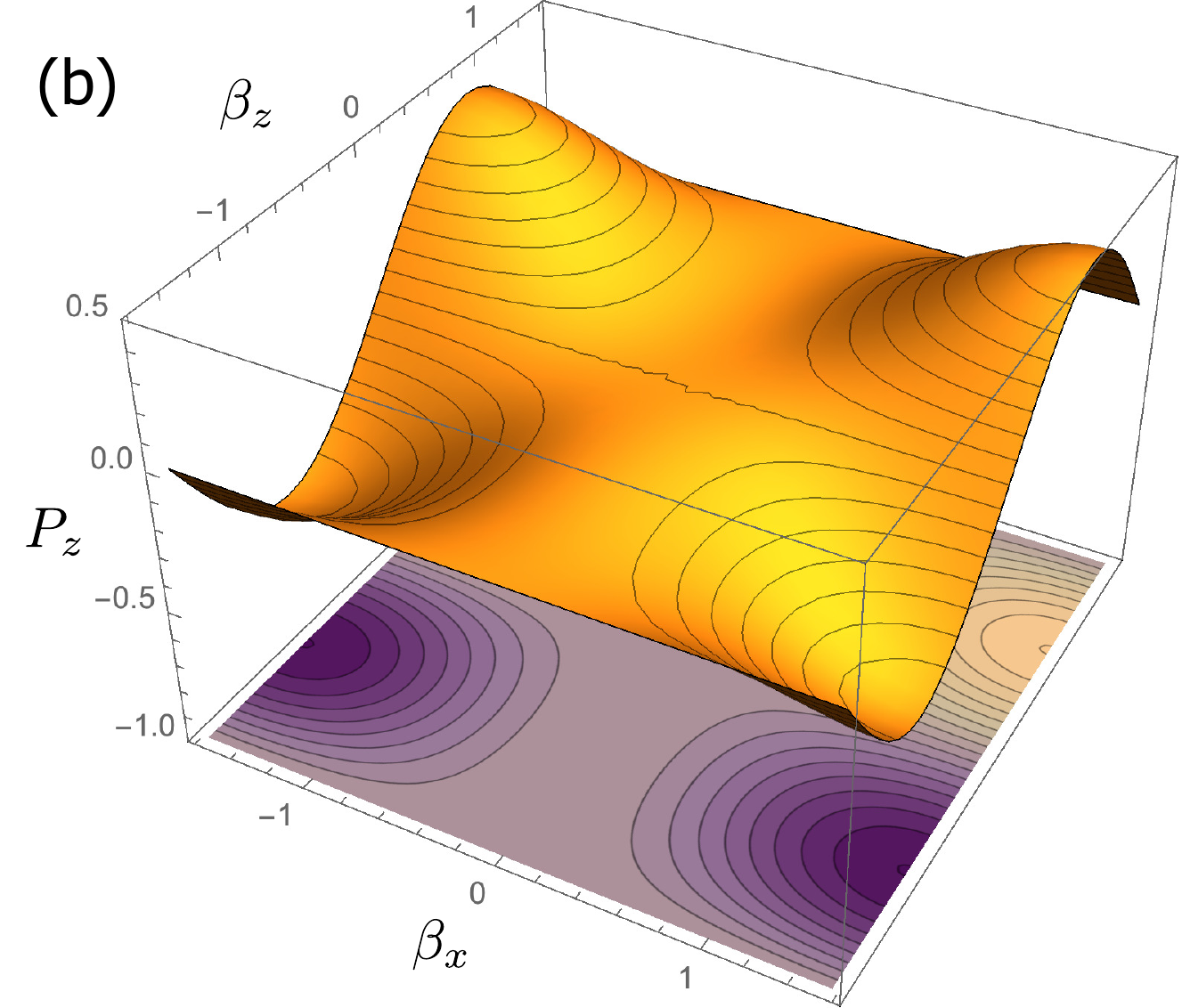}
\caption{(Color online) \footnotesize   
(a) Plot of the universal function $\mathcal{G}(\Phi;k)$ \eqref{eqn:universal-fn-square} that determines $g$ 
through the ACP $\lambda_{\text{AC}}(\beta_x,\beta_z)$.  
(b) 3D plot of the $z$ component of the transmitted electron spin polarization through the square interferometer $P^{\text{(T)}}_z(k=0.7,\beta_x,\beta_z)$  
as a function of $\beta_x$ and $\beta_z$ (unpolarized incoming current is assumed).   Note that, in contrast to $g$ that depends on the traceful part of the ACPF, the spin polarization $P^T_z$ depends 
{\em independly} on $\beta_x$ and $\beta_z$, but it reveals us an information on 
the traceless part of the ACPF.   In addition, it is remarkable to note that even in this simple model the spin polarization reaches about 40\%. 
}
\label{PhaseSquarea}
\end{figure}

With the $4 \times 4$ transmission and reflection matrices at hand (whose calculation is 
detailed in Appendix \ref{sec:solution-square}), we will now inspect the relation between $\lambda_{\text{AC}}$ and the two most accessible  experimental observables, namely,   
 the conductance $g$ and the transmitted spin polarization vector ${\bf P}^{\text{(T)}}$, defined by:
\begin{equation} 
\label{gP}
\begin{split}
& g(k;\beta_x,\beta_y) 
=\mbox{Tr} [t^\dagger t ],  \\
& \mathbf{P}^{\text{(T)}}(k;\beta_x,\beta_y) 
=\frac{\mbox{Tr} [t^\dagger {\bm \Sigma} t ]}{g}   \ \ 
\left(  {\bm \Sigma}={\bf 1}_{2 \times 2} \otimes {\bm \sigma} \right) \; . 
\end{split}
\end{equation} 
\begin{widetext}
The closed expression for the conductance $g$ has the form,
\begin{equation}
g= \mathcal{G}_{\square}\left( \cos \lambda_{\text{AC}}(\beta_x,\beta_z) ;k \right) 
\label{eqn:gPz-1}
\end{equation}
with the ``universal" function $\mathcal{G}_{\square}$ given by:
\begin{equation}
\mathcal{G}_{\square} (\Phi;k) =\frac{16 \sin^2 k[5-2(1-\Phi) \cos 2k]}
{17-8 \cos 2k -4[4 \cos 2k-\cos 4k](1- \Phi)+4(1- \Phi)^2} \; .
\label{eqn:universal-fn-square}
\end{equation}
Clearly, the conductance $g$ depends on the SOC parameters 
{\it only} through $\Phi \equiv \cos \lambda_{\text{AC}}(\beta_x,\beta_z) $  [given in  Eq.~\eqref{PhaseSquare1}].  

For the transmitted spin polarization vector ${\bf P}^{\text{(T)}}$, we get $P^{\text{(T)}}_x=P^{\text{(T)}}_y=0$ and 
\begin{equation}   
 P_z^{\text{(T)}}
 =\frac{2 \sin 2k \sin^2 \beta_x \sin 2 \beta_z}{5-2 \cos 2k(1-\cos  \lambda_{\text{AC}})} 
 =\frac{2 \sin 2k}{5-2 \cos 2k(1-\cos  \lambda_{\text{AC}})}
 \overbrace{\sin \lambda_{\text{AC}} b_z}^{-\tfrac{1}{2} i\, \text{Tr}[\sigma_z F_{\text{AC}}]}. 
 \label{eqn:gPz-2}
 \end{equation}
 where we have used Eq.~\eqref{tracelesssquare}.   
\end{widetext}
\ \\
The fact (proved analytically) that $g$ is a universal function of $\cos \lambda_{\text{AC}}$ (related to the traceful part of $F_{\text{AC}}$) corroborates our earlier results (pertaining to the interferometer under an inhomogeneous electric field) and extends it to multichannel devices. 
It is displayed graphically in Fig.~\ref{PhaseSquarea2}.
In Figs.~\ref{PhaseSquarea2} (a) and (b), the ACP $\cos \lambda_{\text{AC}}(\beta_x,\beta_z)$ (together with the plane
$\cos \lambda_{\text{AC}}(\beta_x,\beta_z)=1/2$) and 
a 3D plot of the conductance $g(k;\beta_x,\beta_z)$ for a fixed value of $k$ are shown, respectively.   
From a glance at these two plots, it is evident that 
$g(\beta_x,\beta_z)$ and $\cos  \lambda_{\text{AC}}(\beta_x,\beta_z)$ share the same ``equipotential" lines in common  
implying $g(\beta_x,\beta_z)=\mathcal{G}_{\square} (\Phi;k)$.  
A 3D plot of the universal function $\mathcal{G}_{\square} (\Phi;k)$ is presented in Fig.~\ref{PhaseSquarea}(a). 

 On the other hand, the expression (\ref{eqn:gPz-2}) for  the transmitted spin polarization $P^{\text{(T)}}_z(\beta_x,\beta_z)$ indicates 
that it depends on {\it both} traceful and traceless parts of the ACPF.   
 To the best of our knowledge, a direct relation between an observable quantity and the traceless part 
 of the ACPF has not yet been derived previously. 
We have thus established a direct method to access the {\em full} ACPF via mesoscopic interferometry.   
Moreover, as shown in Fig.~\ref{PhaseSquarea}(b), the degree of spin polarization is quite sizable even for such a simple devise.    
   
\section{Summary}
\label{summary}
Let us now briefly summarize our main results.  
Starting from the SU(2)-invariant formulation of the Schr\"odinger equation\cite{Anandan-89}\cite{Frohlich-S-93} in its dimensionless form (\ref{dlse}), 
we explained, in Eq.~\eqref{eqn:Wilson-line-discretized}, the construction of the 
SU(2) gauge transformation $F_{\text{AC}}(\theta;\beta)$ that locally eliminates the SU(2) vector potential 
[that takes its value in the Lie algebra su(2)]. 
It is stressed that the matrix-valued Aharonov-Casher phase factor (ACPF)  
 constructed in Eq.~\eqref{ACPF} 
is acquired by an electron subject to SOC,  that moves adiabatically along a closed curve. The ACPF is an SU(2) matrix that is decomposed 
into its traceful and traceless parts, and each part has its own physical content. 

The relation between the Aharonov-Casher phase and the non-Abelian phase factor is 
then clarified.  
Strictly  speaking, the Aharonov-Casher phase should be identified with a $2 \times 2$ matrix $\varphi_{\text{AC}}\in \text{su(2)}$ 
that appears in the exponent of the SU(2) matrix phase factor and carries the {\em full} information on the interference effects, 
as explained in relation to Eq.~(\ref{phiAC}).  Yet, in some circumstances, the
single gauge-invariant quantity $\lambda_{\text{AC}}$ derived only from the traceful part of the full non-Abelian phase factor may also serve as  
the Aharonov-Casher phase. 
In particular, once the geometry of the interferometer is given, $\cos \lambda_{\text{AC}}$ defined in Eq.~\eqref{eqn:ACP-vs-F} 
completely determines the dimensionless conductance $g$ and is experimentally relevant.   

To clarify the relevance of the Aharonov-Casher phase factor to experiments in mesoscopic systems, 
the problem of electron transport through various interferometers subject to SOC is addressed.   
First, the scattering problem of electrons transmitting through a one-dimensional ring interferometer subject to Rashba SOC 
induced by a uniform perpendicular electric field of constant strength $\beta$ is analyzed.  
The gauge transformation $F_{\text{AC}}(\theta;\beta)$ that locally removes the SU(2) vector potential is given in Eq.(\ref{eqn:F-theta-beta}), 
from which the phase factor was deduced as $F_{\text{AC}}(2 \pi;\beta)$.   
The dimensionless conductance $g$ is given explicitly in Eq.~\eqref{eqn:gring} and is shown to be a simple rational function of 
the traceful part $\cos \lambda_{\text{AC}}$ [Eq.~\eqref{eqn:coslambda1}] of $F_{\text{AC}}(2 \pi;\beta)$.  

In order to show that the relation between $\cos \lambda_{\text{AC}}$ and the conductance $g$ is more general, 
we considered a similar scattering problem under an {\it inhomogeneous} perpendicular electric field (where the corresponding Rashba SOC 
is specified by {\it any number of dimensionless SOC parameters}).  
The dimensionless conductance $g$ has precisely the same functional-dependence on 
$\cos \lambda_{\text{AC}}$ as in the previous case [see Eqs.~\eqref{eqn:gring} and \eqref{eqn:expression-g-inhom}], 
with the appropriate replacement of $\cos \lambda_{\text{AC}}$ 
[see, for example, Eq.~\eqref{lambdab1b2} for the expression of 
$\cos \lambda_{\text{AC}}$  in the case of two strength parameters $\beta_1,\beta_2$].  
This central result 
unambiguously confirms our statement that the conductance $g$ depends on the SOC 
parameters {\it only} through the traceful part $\cos \lambda_{\text{AC}}$ of the Aharonov-Casher phase factor.   
We again stress that this remarkable universality is not at all obvious from the SU(2) gauge invariance alone. 

In order to experimentally access the traceless part of the ACPF, 
it is necessary to consider the electron spin polarization. We briefly explained in Sec.~\ref{sec:NoP} the reasons 
for the absence of electron spin polarization in 
interferometers with strictly two one-dimensional leads (the source and drain) whose Hamiltonian is time-reversal invariant.  
In order to get some hints about possible relations between the electron spin polarization and the Aharonov-Casher phase factor,  
a tight-binding model of square interferometer with 
two incoming and two outgoing channels was studied (see Fig.~\ref{SquareScatt}), 
for which both the conductance {\it and} electron spin polarization were calculated.  
We solved the scattering problem for this square interferometer in the presence of an {\it inhomogeneous} perpendicular electric field 
that generates Rashba SOC with two different dimensionless parameters $\beta_x$ and $\beta_z$.   
As shown in Eq.~(\ref{eigenvalues-square}), the energies of an electron hopping on the square (which is detached from the rest of the system) are 
given by simple trigonometric functions of $\lambda_{\text{AC}}$.  Moreover, as shown in 
Eq.~\eqref{eqn:gPz-1}, the conductance is again a simple rational function of $\cos \lambda_{\text{AC}}$, 
that is the traceful part of the phase factor.  
On the other hand, electron spin polarization is related in Eq.~(\ref{eqn:gPz-2}) both to the traceful part (through $\cos \lambda_{\text{AC}}$) 
and to the traceless part (through  $\sin \lambda_{\text{AC}} \widehat{\bf b}$) of the phase factor. 
Thus, the elusive Aharonov-Casher phase factor in its most general structure is experimentally accessible with mesoscopic interferometry. 
 
\begin{acknowledgments}
Y.A. is grateful to the hospitality of Yukawa Institute for Theoretical Physics where this work has been initiated. His research is supported in part by the Israeli Science Foundation  (grant 400/12) and by the NYU-Shanghai university research fund.
He also benefitted from discussions with Pier Mello, Y. B. Band and D. Ariad.  
Discussions with F. Pi\`{e}chon are highly appreciated. 
N.N. was supported by JST CREST Grant Number JPMJCR1874 and JPMJCR16F1, Japan,  
and JSPS KAKENHI Grant No.~18H03676 and 26103006.
K.T. is supported in part by JSPS KAKENHI Grants No.~15K05211 and No.~18K03455.  
This project was supported in part by JSPS and ISF under Japan-Israel Research Cooperative Program.
\end{acknowledgments}
\appendix
\section{Calculation of phase factors for ring interferometers}
\label{app-Ring}
In this appendix, we describe how to calculate path-ordered ($\mathcal{P}$) product (Wilson line) of SU(2) matrices 
that appear in Secs.~\ref{sec:Ring} and \ref{sec:Ringinhom}. Calculations are carried out both for 
the Rashba and the Dresselhaus SOC schemes. 
\subsection{Case with a single spin-orbit coupling, discussed in section \ref{sec:Ring}}
\label{sec:F-for-single-beta}
Recall the path-ordered integral [see Eqs.~\eqref{eqn:Wilson-line-discretized}], 
\begin{equation}
\begin{split}
& \psi(\theta) = {\cal P}\int_0^\theta  \text{e}^{i \beta(\theta') \widehat{\bf n}(\theta') \cdot {\bm \sigma} d \theta'} \psi(0)  \\
& =\lim_{N \to \infty} \left\{ \prod_{\substack{n=1\\ \longleftarrow }}^{N}
\text{e}^{ i \beta(n \Delta \theta) \widehat{\bf n}(n \Delta \theta) {\cdot} {\bm \sigma} 
\Delta \theta  } \right \} \psi(0) 
\equiv F_{\text{AC}}(\theta ; \beta) \psi(0) \; , 
\end{split}
\label{eqn:wave-fn-by-ACPF} 
\end{equation}
with 
$\boldsymbol{\sigma} =(\sigma_{x},\sigma_{y},\sigma_{z})$ and $\Delta \theta=\frac{\theta}{N}$.  
For the ring interferometer considered in Fig.~\ref{RingInter}(a), we take $\hat{\mathbf{n}}(\theta)=(\cos\theta,\sin \theta,0)$ 
for the Rashba spin-orbit interaction and $\hat{\mathbf{n}}(\theta)=(\sin \theta,\cos \theta,0)$ for the Dresselhaus spin-orbit 
interaction.  
\subsubsection{Rashba spin-orbit interaction}
We begin with the case with the Rashba spin-orbit interaction where the unit vector $\hat{\bf n}(\theta)$ is 
pointing in the radial direction: $\hat{\mathbf{n}}(\theta) = (\cos \theta,\sin \theta,0)$.   
By definition, $F_{\text{AC}}(\theta+\Delta\theta;\beta)$ is obtained by multiplying $F_{\text{AC}}(\theta;\beta)$ by 
$\be^{i \beta \Delta \theta \hat{\mathbf{n}}(\theta){\cdot}\boldsymbol{\sigma}}$ on the left:
\begin{equation}
\begin{split}
& F_{\text{AC}}(\theta+\Delta\theta;\beta) = \be^{i \beta \, \hat{\mathbf{n}}(\theta){\cdot}\boldsymbol{\sigma} \Delta \theta} F_{\text{AC}}(\theta;\beta) \\
& \approx \left\{ \mathbf{1}_{2 \times 2} 
+ i \beta \Delta \theta\, \hat{\mathbf{n}} (\theta){\cdot}\boldsymbol{\sigma} \right\}  F_{\text{AC}}(\theta;\beta)
+ \text{O}(\Delta\theta^{2})  \\
& \approx \left\{ \mathbf{1}_{2 \times 2} + i \beta \Delta \theta 
\begin{pmatrix}
0 & \be^{-i\theta} \\
\be^{i\theta} & 0 
\end{pmatrix}
 \right\}  F_{\text{AC}}(\theta;\beta)
+ \text{O}(\Delta\theta^{2}) \; .
\end{split}
\end{equation}
This procedure enables derivation of a differential equation for $F_{\text{AC}}(\theta;\beta)$:
\begin{equation}
\frac{d}{d\theta} F_{\text{AC}}(\theta;\beta) =  i \beta  
\begin{pmatrix}
0 & \be^{-i\theta} \\
\be^{i\theta} & 0 
\end{pmatrix}
F_{\text{AC}}(\theta;\beta) 
= \left\{ i \beta \hat{\mathbf{n}}(\theta){\cdot}\boldsymbol{\sigma} \right\} F_{\text{AC}}(\theta; \beta)   \; .
\label{eqn:diff-eq-AC-phase}
\end{equation}
The solution to this set equations takes the following form:
\begin{equation}
\begin{split}
F_{\text{AC}}(\theta;\beta) &= 
\begin{pmatrix}
F_{11}(\theta;\beta) & F_{12}(\theta;\beta)  \\
F_{21}(\theta;\beta)  & F_{22}(\theta;\beta) 
\end{pmatrix}  
\\
& = \mathcal{G}_{\beta}(\theta)
\begin{pmatrix}
C_1 & C_3 \\ C_2 & C_4 
\end{pmatrix}
= \mathcal{G}_{\beta}(\theta)\mathbf{C} \; , 
\end{split}
\end{equation}
where  the $\theta$-dependent part is defined as
\begin{equation}
\begin{split}
& \mathcal{G}_{\beta}(\theta) \\
& =
\begin{pmatrix}
\be^{-\frac{i}{2}  \left(1+\sqrt{4 \beta ^2+1} \right)\theta } & \be^{-\frac{i}{2} \left(1- \sqrt{4 \beta ^2+1} \right)\theta } \\
- \frac{1 + \sqrt{4 \beta ^2+1}}{2\beta }  \be^{\frac{i}{2} \left(1- \sqrt{4 \beta ^2+1} \right)\theta } 
& - \frac{1 - \sqrt{4 \beta ^2+1}}{2\beta }  \be^{\frac{i}{2} \left(1 + \sqrt{4 \beta ^2+1} \right)\theta } 
\end{pmatrix} 
\\
& \mathcal{G}_{\beta}(-\theta) = \mathcal{G}_{\beta}^{\ast}(\theta)
\end{split}
\label{eqn:expression-G}
\end{equation}
and the constant matrix $\mathbf{C}$ is determined by the initial condition:
\begin{equation}
\mathcal{G}_{\beta}(\theta=0) \mathbf{C} = F_{\text{AC}}(\theta=0;\beta) \; (= \text{a given matrix} ) \; .
\end{equation}
\begin{widetext}

Physically, we need the solution satisfying the initial condition $F_{\text{AC}}(\theta=0;\beta) = \mathbf{1}_{2 \times 2}$, that is, 
\begin{equation}
\begin{split}
& F_{\text{AC}}(\theta;\beta) =  \mathcal{G}_{\beta}(\theta) \mathcal{G}_{\beta}^{-1}(0) \\
&= 
\begin{pmatrix}
\frac{\be^{-\frac{i \theta }{2}} \left\{  \sqrt{4 \beta ^2+1} \cos \left(\frac{1}{2} \sqrt{4 \beta ^2+1} \theta
   \right)+i \sin \left(\frac{1}{2} \sqrt{4 \beta ^2+1} \theta \right)\right\}}{\sqrt{4 \beta ^2+1}} & \frac{2 i \, 
   \be^{-\frac{i \theta }{2}} \beta  \sin \left(\frac{1}{2} \sqrt{4 \beta ^2+1} \theta \right)}{\sqrt{4 \beta^2+1}} \\
\frac{2 i \, \be^{\frac{i \theta }{2}} \beta  \sin \left(\frac{1}{2} \sqrt{4 \beta ^2+1} \theta \right)}{\sqrt{4\beta ^2+1}} 
& \frac{\be^{\frac{i \theta }{2}} \left\{ \sqrt{4 \beta ^2+1} \cos \left(\frac{1}{2} \sqrt{4 \beta^2+1} \theta\right)   
 -i \sin \left(\frac{1}{2} \sqrt{4 \beta ^2+1} \theta \right)\right\} }{\sqrt{4 \beta ^2+1}}  
\end{pmatrix}  \\
 &= 
 \be^{-\frac{i \theta }{2}\sigma_{z} }  
\exp\left\{ 
i \theta y(\beta) \boldsymbol{N}^{\text{(R)}} (\beta) {\cdot} \boldsymbol{\sigma} \right\}   \; , 
\end{split}
\label{eqn:F-theta-beta-formula}
\end{equation}
where the unit vector $\boldsymbol{N}^{\text{(R)}} (\beta)$ is defined by
\begin{equation}
\boldsymbol{N}^{\text{(R)}} (\beta) = (\sin \gamma(\beta),0,\cos\gamma(\beta)) \equiv 
\left(  \frac{2 \beta }{\sqrt{4 \beta^2+1}} ,0,  \frac{1}{\sqrt{4 \beta ^2+1}} \right)   \; .
\end{equation}
Together with Eq.~\eqref{eqn:expression-G}, we can easily verify the following properties:
\begin{equation}
\begin{split}
& F_{\text{AC}}(-\theta; \beta) =  \mathcal{G}_{\beta}(-\theta)\mathcal{G}_{\beta}^{-1}(0) 
=  \mathcal{G}^{\ast}_{\beta}(\theta)\mathcal{G}_{\beta}^{-1}(0) = F_{\text{AC}}^{\ast} (\theta; \beta)  \\
& F_{\text{AC}}^{\dagger}(\theta; \beta) F_{\text{AC}}(\theta; \beta) =
F_{\text{AC}}^{\text{T}}(-\theta; \beta) F_{\text{AC}}(\theta; \beta) = \mathbf{1} \\ 
& \quad \Leftrightarrow \; 
F_{\text{AC}}^{-1}(\theta; \beta) = F_{\text{AC}}^{\dagger}(\theta; \beta) = F_{\text{AC}}^{\text{T}}(-\theta; \beta)
\quad (\text{unitary}) \; .
\end{split}
\label{eqn:F-non-Abelian-AC}
\end{equation}
Using the two eigenvectors 
\begin{equation}
\boldsymbol{u}_{+}(\beta)
=\begin{pmatrix} \cos \frac{\gamma(\beta)}{2} \\ \sin \frac{\gamma(\beta)}{2} \end{pmatrix} \; , \;\; 
\boldsymbol{u}_{-}(\beta) = 
\begin{pmatrix} - \sin \frac{\gamma(\beta)}{2} \\ \cos \frac{\gamma(\beta)}{2} \end{pmatrix}
\end{equation}
of $\boldsymbol{N}^{\text{(R)}} (\beta) {\cdot} \boldsymbol{\sigma}$, we can obtain the explicit expressions for the wave function 
\eqref{eqn:wave-fn-by-ACPF} satisfying the periodic boundary condition as:
\begin{equation}
\begin{split}
& \psi^{(n)}_{\pm} (\theta) = \frac{1}{\sqrt{2\pi}} \be^{i k_{\pm}^{(n)} \theta} F_{\text{AC}}(\theta;\beta)  \boldsymbol{u}_{\pm}(\beta) 
= \frac{1}{\sqrt{2\pi}}  \be^{i n \theta} \be^{\mp \frac{i}{2} \theta (1\pm \sigma_{z})} \boldsymbol{u}_{\pm}(\beta)  \\
& \phantom{\Psi^{(n)}_{\pm} (\theta)}  = 
\frac{1}{\sqrt{2\pi}} \be^{i n \theta} \begin{pmatrix} \be^{-i \theta} \cos \frac{\gamma(\beta)}{2} \\ \sin \frac{\gamma(\beta)}{2} \end{pmatrix} 
\;\; \text{(for ``$+$'')} \; , \quad 
\frac{1}{\sqrt{2\pi}} \be^{i n \theta} \begin{pmatrix} - \sin \frac{\gamma(\beta)}{2} \\ \be^{i \theta} \cos \frac{\gamma(\beta)}{2} \end{pmatrix}  
\;\; \text{(for ``$-$'')}  \\
& k_{\pm}^{(n)} = n \mp \frac{1}{2\pi} \lambda_{\text{AC}}(\beta) 
= n \mp \frac{1}{2} \left(1 + \sqrt{4 \beta ^2+1} \right) \quad (n \in \mathbb{Z})   \\
& \veps_{\pm}^{(n)} =\frac{2 m R^2}{\hbar^2} E_{\pm}^{(n)} = \left( k_{\pm}^{(n)} \right)^2   \; .
\end{split}
\end{equation}
In these energy eigenstates, the electron spin is {\em tilted} in the direction of 
$(\cos\theta\sin\gamma(\beta),\sin\theta\sin\gamma(\beta),\cos \gamma(\beta))$.   

From Eq.~\eqref{eqn:F-theta-beta-formula}, we obtain the non-Abelian ACPF:\cite{Qian-S-94}
\begin{equation}
F_{\text{AC}}(\theta=2\pi;\beta) =  -  \exp\left\{ 
2\pi i y(\beta) \boldsymbol{N}^{\text{(R)}} (\beta) {\cdot} \boldsymbol{\sigma} 
\right\}  \; , 
\end{equation}
whose eigenvalue $- \be^{ \pm i \pi \sqrt{4 \beta ^2+1}}$ consists of two parts:
\begin{equation}
- \be^{\pm i \pi \sqrt{4 \beta ^2+1}} = \be^{ \pm i 2\pi  \beta \sin \Theta(\beta)} \be^{\pm i \pi (1+\cos\Theta(\beta)) }   \; .
\end{equation}
The first part is the usual ACP \eqref{eqn:usual-ACP} due to the projection $\beta \sin \Theta(\beta)$ of the effective magnetic field 
in the direction of the polarized spin, while the second may be interpreted as the spin Berry phase coming from the spinorial  
part of the wave function $\psi^{(n)}_{\pm} (\theta)$, thus reproducing the observation in Ref.~\citen{Qian-S-94}.  
The Aharonov-Casher phase Eq.~\eqref{eqn:coslambda1} is given by the gauge-invariant trace of $F_{\text{AC}}(\theta=2\pi;\beta)$:
\begin{equation}
\cos \lambda_{\text{AC}}(\beta) = 
\frac{1}{2} \text{Tr} \, F_{\text{AC}}(\theta=2\pi;\beta) = -\cos \left(\pi  \sqrt{4 \beta ^2+1}\right) \; .
\label{eqn:AC-phase-uniform-beta}
\end{equation}
Using the explicit form \eqref{eqn:F-theta-beta-formula}, we can show the following relation:
\begin{equation}
F_{\text{AC}}(2 \pi ;\beta ) + F_{\text{AC}}^{-1}(2 \pi ;\beta ) = 2 \cos \lambda_{\text{AC}}(\beta) \mathbf{1}_{2{\times}2} \; .
\label{eqn:identity-F-Finv}
\end{equation}
\end{widetext}

\begin{figure}[htb]
\begin{center}
\includegraphics[width=0.7\columnwidth]{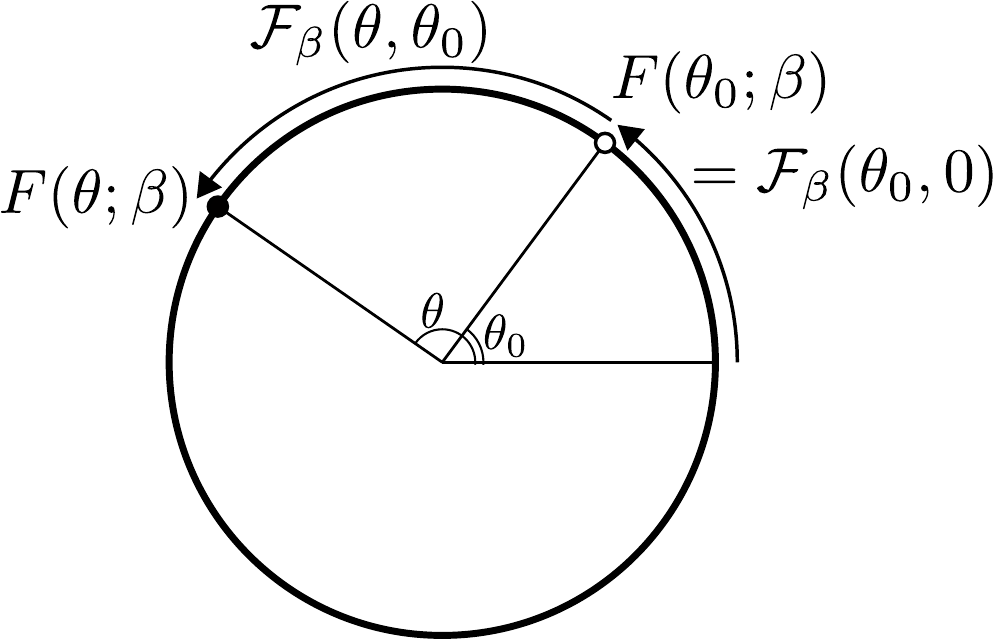}
\end{center}
\caption{Composition of two Wilson lines to obtain $F_{\text{AC}}(\theta;\beta)$.  
\label{fig:composition-Wilson-line}}
\end{figure}

For later use, it is convenient to consider the partial phase acquired between $\theta_0$ and $\theta$: 
$\mathcal{F}_{\beta}(\theta,\theta_0)=\mathcal{G}_{\beta}(\theta)\mathbf{C}_{\theta_0}$ 
that satisfies a slightly different initial condition (Fig.~\ref{fig:composition-Wilson-line}):
\begin{equation}
\begin{split}
& \mathcal{F}_{\beta}(\theta_0,\theta_0) = \mathcal{G}_{\beta}(\theta_0)\mathbf{C}_{\theta_0} = \mathbf{1}_{2 \times 2}   \\
& \quad \Rightarrow \quad 
\mathcal{F}_{\beta}(\theta,\theta_0)=\mathcal{G}_{\beta}(\theta) \mathcal{G}^{-1}_{\beta}(\theta_0)
\end{split}
\label{eqn:def-cal-F}
\end{equation}
[$F_{\text{AC}}(\theta;\beta)$ is a special case of $\mathcal{F}_{\beta}(\theta,\theta_0)$ with $\theta_0 = 0$; 
$F_{\text{AC}}(\theta;\beta)=\mathcal{F}_{\beta}(\theta,0)$].  
It is easy to see that $\mathcal{F}_{\beta}(\theta,\theta_0)$ measures the non-Abelian phase accumulated between $\theta_{0}$ and $\theta$:
\begin{equation}
\mathcal{F}_{\beta}(\theta,\theta_0)
= \mathcal{P} \exp \left\{ 
i \beta \int_{\theta_0}^{\theta} \!  \hat{\mathbf{n}} (\theta^{\prime}) {\cdot} \boldsymbol{\sigma} 
d\theta^{\prime}   \right\}  \; .
\label{eqn:F-propagator-path-ordered}
\end{equation}   
It is important to note that $\mathcal{F}_{\beta}(\theta,\theta_0)$ cannot be written as a function of $\theta-\theta_0$ 
since the differential equation \eqref{eqn:diff-eq-AC-phase} depends {\em explicitly} 
on $\theta$ itself [see Eq.~\eqref{eqn:F-propagator}].  
Rather, $\mathcal{F}_{\beta}(\theta,\theta_0)$ depends on $\theta$ and $\theta_0$ separately as
\begin{equation}
\begin{split}
\mathcal{F}_{\beta}(\theta,\theta_0) =&  \mathcal{G}_{\beta}(\theta) \mathcal{G}_{\beta}^{-1}(\theta_0) 
= \left\{ \mathcal{G}_{\beta}(\theta) \mathcal{G}^{-1}(0) \right\} \left\{ \mathcal{G}_{\beta}(\theta_0) \mathcal{G}^{-1}(0) \right\}^{-1}  \\ 
=&  F_{\text{AC}}(\theta; \beta) F_{\text{AC}}^{-1}(\theta_0; \beta) \; .
\end{split}
\label{eqn:F-propagator}
\end{equation}
Obviously, $\mathcal{F}_{\beta}(\theta,\theta_0)$ satisfies the following properties:
\begin{equation}
\begin{split}
& \mathcal{F}_{\beta}(\theta_2, \theta_1) = \mathcal{F}_{\beta}^{-1} (\theta_1, \theta_2)  \\
& \mathcal{F}_{\beta}^{-1} (\theta_1, \theta_2)  
= \mathcal{F}^{\dagger}_{\beta}(\theta_1, \theta_2)  \; , \;\; 
\text{det} \, \mathcal{F}_{\beta}(\theta_1, \theta_2) =1 \\ 
& [ \mathcal{F}_{\beta}(\theta_1, \theta_2) \in \text{SU(2)} ] \;, 
\end{split}
\label{eqn:F-transposition}
\end{equation}
which immediately results from 
$\mathcal{F}_{\beta}(\theta_1,\theta_2)\mathcal{F}_{\beta}(\theta_2,\theta_1)
=\mathcal{F}_{\beta}(\theta_1,\theta_1)=\mathbf{1}_{2 \times 2}$.   
It is easy to verify the following composition law [$(0 \to \theta_0) \times (\theta_0 \to \theta)=(0 \to \theta)$] 
(see Fig.~\ref{fig:composition-Wilson-line}):
\begin{equation}
\begin{split}
& \underbrace{\mathcal{F}_{\beta}(\theta,\theta_0)}_{\theta \leftarrow \theta_0}
F_{\text{AC}}(\theta_0;\beta) = \mathcal{F}_{\beta}(\theta,\theta_0) \mathcal{F}_{\beta}(\theta_0, 0)  \\
& = \left\{\mathcal{G}_{\beta}(\theta)\mathcal{G}^{-1}(\theta_0) \right\}
\left\{ \mathcal{G}_{\beta}(\theta_0)\mathcal{G}^{-1}(0)  \right\} \\
&=  \mathcal{G}_{\beta}(\theta) \mathcal{G}_{\beta}^{-1}(0)  = F_{\text{AC}}(\theta;\beta) \; .
\end{split}
\label{eqn:F-composition-rule}
\end{equation}

Another useful property is the shift invariance of $\mathcal{F}_{\beta}(\theta,\theta_0)$ resulting from 
the invariance of the integrand of Eq.~\eqref{eqn:F-propagator-path-ordered} under $\theta \to \theta+2\pi$, 
$\theta_0 \to \theta_0 + 2\pi$.  
In fact, noting that the shift $\theta \to \theta + 2\pi$ amounts to multiplication of a constant ($\theta$-independent) 
matrix on the right,
\begin{equation}
\mathcal{G}_{\beta}(\theta+2\pi) = \mathcal{G}_{\beta}(\theta) 
\begin{pmatrix}
-\be^{-i \pi  \sqrt{4 \beta ^2+1}} & 0 \\
 0 & -\be^{i \pi  \sqrt{4 \beta ^2+1}}
\end{pmatrix}  
\; ,
\label{eqn:periodicity-G}
\end{equation}
one immediately verifies the $2\pi$-periodicity of $\mathcal{F}_{\beta}(\theta ;\theta_0)$: 
\begin{equation}
\begin{split}
& \mathcal{F}_{\beta}(\theta+2\pi;\theta_0+2\pi) = \mathcal{G}_{\beta}(\theta+2\pi)\mathcal{G}_{\beta}^{-1}(\theta_0 +2\pi)  \\
& = \mathcal{G}_{\beta}(\theta)\mathcal{G}_{\beta}^{-1}(\theta_0)  
= \mathcal{F}_{\beta}(\theta ;\theta_0)  \; .
\end{split}
\label{eqn:F-2pi-shift}
\end{equation}
Another useful property is a $\pi$ shift,
\begin{equation}
\begin{split}
& \mathcal{F}_{\beta}(\theta+\pi, \theta_0+\pi) =  \mathcal{P} \exp \left\{ 
i \beta \int_{\theta_0+\pi}^{\theta+\pi} \!  \hat{\mathbf{n}}(\theta^{\prime}) {\cdot} \boldsymbol{\sigma} 
d\theta^{\prime}   \right\}   \\
& = \mathcal{P} \exp \left\{ 
- i \beta \int_{\theta_0}^{\theta} \!  \hat{\mathbf{n}}(\xi) {\cdot} \boldsymbol{\sigma} 
d\xi   \right\}  = \mathcal{F}_{-\beta}(\theta, \theta_0)  \; .
\end{split}
\label{eqn:F-pi-shift}
\end{equation}
\subsubsection{Dresselhaus spin-orbit interaction}
We can follow the same steps to derive the path-ordered product of the SU(2) phase $F_{\text{AC}}^{\text{(D)}}(\theta;\beta)$ 
and the AC-phase for the same interferometer with 
the Dresselhaus spin-orbit interaction where $\hat{\mathbf{n}}(\theta)=(\sin \theta,\cos \theta,0)$.   
The differential equation for $F^{\text{(D)}}(\theta;\beta)$ now reads as:  
\begin{equation}
\frac{d}{d\theta} F_{\text{AC}}^{\text{(D)}}(\theta;\beta) =  i \beta  
\begin{pmatrix}
0 & -i \be^{i\theta} \\
i \be^{-i\theta} & 0 
\end{pmatrix}
F_{\text{AC}}^{\text{(D)}} (\theta;\beta) \; .
\label{eqn:diff-eq-AC-phase-2}
\end{equation}
\begin{widetext}
Comparing this with \eqref{eqn:diff-eq-AC-phase} and using
\begin{equation*}
\begin{pmatrix}
0 & -i \be^{i\theta} \\
i \be^{-i\theta} & 0 
\end{pmatrix} 
= \be^{-i\frac{\pi}{4}\sigma_{z}} 
\begin{pmatrix}
0 & \be^{+i\theta} \\
\be^{- i\theta} & 0 
\end{pmatrix}
\be^{+i\frac{\pi}{4}\sigma_{z}} \; , 
\end{equation*}
one sees that the solution is given by
\begin{equation}
\begin{split}
& F_{\text{AC}}^{\text{(D)}} (\theta;\beta) = \be^{-i\frac{\pi}{4}\sigma_{z}} F_{\text{AC}} (-\theta; -\beta) \be^{+i\frac{\pi}{4}\sigma_{z}}  \\
&= 
\begin{pmatrix}
\frac{\be^{\frac{i \theta }{2}} \left\{ \sqrt{4 \beta ^2+1} \cos \left(\frac{1}{2} \sqrt{4 \beta ^2+1} \theta \right)
 - i \sin \left(\frac{1}{2} \sqrt{4 \beta^2+1} \theta \right)\right\}}{\sqrt{4 \beta ^2+1}} & 
  \frac{2 \be^{\frac{i \theta }{2}} \beta  \sin \left(\frac{1}{2} \sqrt{4 \beta ^2+1} \theta \right)}{\sqrt{4 \beta^2+1}} \\
 -\frac{2 \be^{-\frac{i \theta }{2}} \beta  \sin \left(\frac{1}{2} \sqrt{4 \beta ^2+1} \theta \right)}{\sqrt{4\beta^2+1}} 
 & \frac{\be^{-\frac{i \theta }{2}} \left\{ \sqrt{4 \beta ^2+1} \cos \left(\frac{1}{2} \sqrt{4 \beta^2+1} \theta \right) 
 + i \sin \left(\frac{1}{2} \sqrt{4 \beta ^2+1} \theta \right)\right\}}{\sqrt{4 \beta^2+1}}  
\end{pmatrix}  \\
&= \be^{\frac{i \theta }{2}\sigma_{z} }  
\left\{ 
\cos \left(\frac{1}{2} \sqrt{4 \beta ^2+1} \theta \right) \mathbf{1} 
-  i \sin \left(\frac{1}{2} \sqrt{4 \beta ^2+1} \theta \right) 
\left( \frac{- 2 \beta }{\sqrt{4 \beta^2+1}} \sigma_{y} + \frac{1}{\sqrt{4 \beta ^2+1}}\sigma_{z} \right)
\right\} \\
&= 
\be^{\frac{i \theta }{2}\sigma_{z} }  
\exp\left\{ 
- i \frac{\theta}{2} \sqrt{4 \beta ^2+1}   
\boldsymbol{N}^{\text{(D)}} (\beta) {\cdot} \boldsymbol{\sigma} \right\}   \; (\in \text{SU(2)})
\end{split}
\label{eqn:F_Dresselhaus}
\end{equation}  
with 
\begin{equation}
\boldsymbol{N}^{\text{(D)}} (\beta) \equiv 
\left(  0, \frac{- 2 \beta }{\sqrt{4 \beta^2+1}} ,  \frac{1}{\sqrt{4 \beta ^2+1}} \right)   \; .
\end{equation}
\end{widetext}
All the properties found for the Rashba spin-orbit interaction [e.g., \eqref{eqn:F-composition-rule}, 
\eqref{eqn:periodicity-G}, and \eqref{eqn:F-2pi-shift}] holds for the Dresselhaus spin-orbital interaction as well.  
From Eq.~\eqref{eqn:F_Dresselhaus}, we can read off the AC-phase, which is exactly the same 
as that for Rashba spin-orbit interaction:
\begin{equation}
\cos \lambda^{\text{(D)}}_{\text{AC}} = 
\frac{1}{2} \text{Tr} \, F_{\text{AC}}^{\text{(D)}}(\theta=2\pi;\beta) = -\cos \left(\pi  \sqrt{4 \beta ^2+1}\right) \; .
\label{eqn:AC-phase-uniform-beta-2}
\end{equation}

\subsection{Case with different spin-orbit couplings}
\label{sec:F-for-two-beta}
Now let us consider the situation shown in Fig.~\ref{RingInter}(b).  
Using the phase $\mathcal{F}_{\beta}(\theta, \theta_0)$ defined above [see Eq.~\eqref{eqn:def-cal-F}], 
the total ``phase'' ${\cal F}(\theta; \beta_1,\beta_2,\alpha)$ 
acquired along the path $\text{P}_{1}\to \text{P}_{2} \to \text{M} \to \text{P}_{1}$ is calculated as [see Fig.~\ref{fig:two-beta}(b)]
\begin{equation}
\begin{split}
& F_{\text{AC}}(\text{P}_{1} \leftarrow \text{M} \leftarrow \text{P}_{2}) F_{\text{AC}}(\text{P}_{2}\leftarrow \text{P}_{1})  \\
&= \mathcal{F}_{\beta_{2}}(2\pi,\alpha) \mathcal{F}_{\beta_{1}}(\alpha;0) 
= \mathcal{G}_{\beta_2}(2\pi) \mathcal{G}^{-1}_{\beta_2}(\alpha) 
\mathcal{G}_{\beta_1}(\alpha) \mathcal{G}^{-1}_{\beta_1}(0)  \\
& = F_{\text{AC}}(2\pi;\beta_2) \left[ F_{\text{AC}}(\alpha;\beta_2)\right]^{-1} F_{\text{AC}}(\alpha;\beta_1)  \\
& \equiv F_{\text{AC}}(2\pi;\alpha;\beta_1,\beta_2)   \; ,
\end{split}
\label{eqn:def-non-Abelian-AB-inhom}
\end{equation}
\begin{widetext}
where we have used Eq.~\eqref{eqn:F-propagator}.   
The trace of the above string of matrices (with $\theta=2\pi$) gives $\lambda_{\text{AC}}$:  
\begin{equation}
\begin{split}
& \cos \lambda_{\text{AC}}(\alpha;\beta_1,\beta_2) 
=\frac{1}{2} \text{Tr} \left\{ \mathcal{F}_{\beta_{2}}(2\pi;\alpha) \mathcal{F}_{\beta_{1}}(\alpha;0) \right\}  \\
& =\frac{1}{2} \text{Tr} \left\{ F_{\text{AC}}(\theta;\beta_2) \left[ F_{\text{AC}}(\alpha;\beta_2)\right]^{-1} F_{\text{AC}}(\alpha;\beta_1) \right\}  
= \frac{1}{2} \text{Tr} \biggl\{
 \mathcal{G}_{\beta_2}(2\pi) \mathcal{G}^{-1}_{\beta_2}(\alpha) 
\mathcal{G}_{\beta_1}(\alpha) \mathcal{G}^{-1}_{\beta_1}(0)    \biggr\}  \\
& =   \biggl\{ 
\frac{4 \beta_{1} \beta_{2}+1}{ \sqrt{4 \beta_{1}^2+1} \sqrt{4 \beta_{2}^2+1}} 
\sin \left(\frac{1}{2} \alpha  \sqrt{4 \beta_{1}^2+1}\right) \sin
   \left(\frac{1}{2} (2 \pi -\alpha ) \sqrt{4 \beta_{2}^2+1}\right)   \\
& \qquad \qquad    - \cos \left(\frac{1}{2} \alpha  \sqrt{4 \beta_{1}^2+1}\right) \cos \left(\frac{1}{2} (2 \pi -\alpha )
   \sqrt{4 \beta_{2}^2+1}\right)  \biggr\} \; .
\end{split}
\label{eqn:AC-phase-two-beta}
 \end{equation}
 As is expected, it is independent of the choice of the starting point.   
 If we set $\beta_1=\beta_2=\beta$, we recover the result \eqref{eqn:AC-phase-uniform-beta}: 
 $\cos \lambda_{\text{AC}}(\alpha;\beta,\beta) = \cos \lambda_{\text{AC}}(\beta)$.  
Being an SU(2) matrix, $F_{\text{AC}}(2\pi;\alpha;\beta_1,\beta_2)$ satisfies the following identity:
\begin{equation}
\begin{split}
F_{\text{AC}}(2\pi;\alpha;\beta_1,\beta_2) + F^{-1}_{\text{AC}}(2\pi;\alpha;\beta_1,\beta_2) 
&= \text{Tr}F_{\text{AC}}(2\pi;\alpha;\beta_1,\beta_2)  \mathbf{1}_{2{\times}2}  \\
&= 2 \cos \lambda_{\text{AC}}(\alpha;\beta_1,\beta_2)  \mathbf{1}_{2{\times}2} \; .
\end{split}
\label{eqn:identity-F-Finv-inhom}
\end{equation}

\begin{figure}[htb]
\begin{center}
\includegraphics[width=1.2\columnwidth,clip]{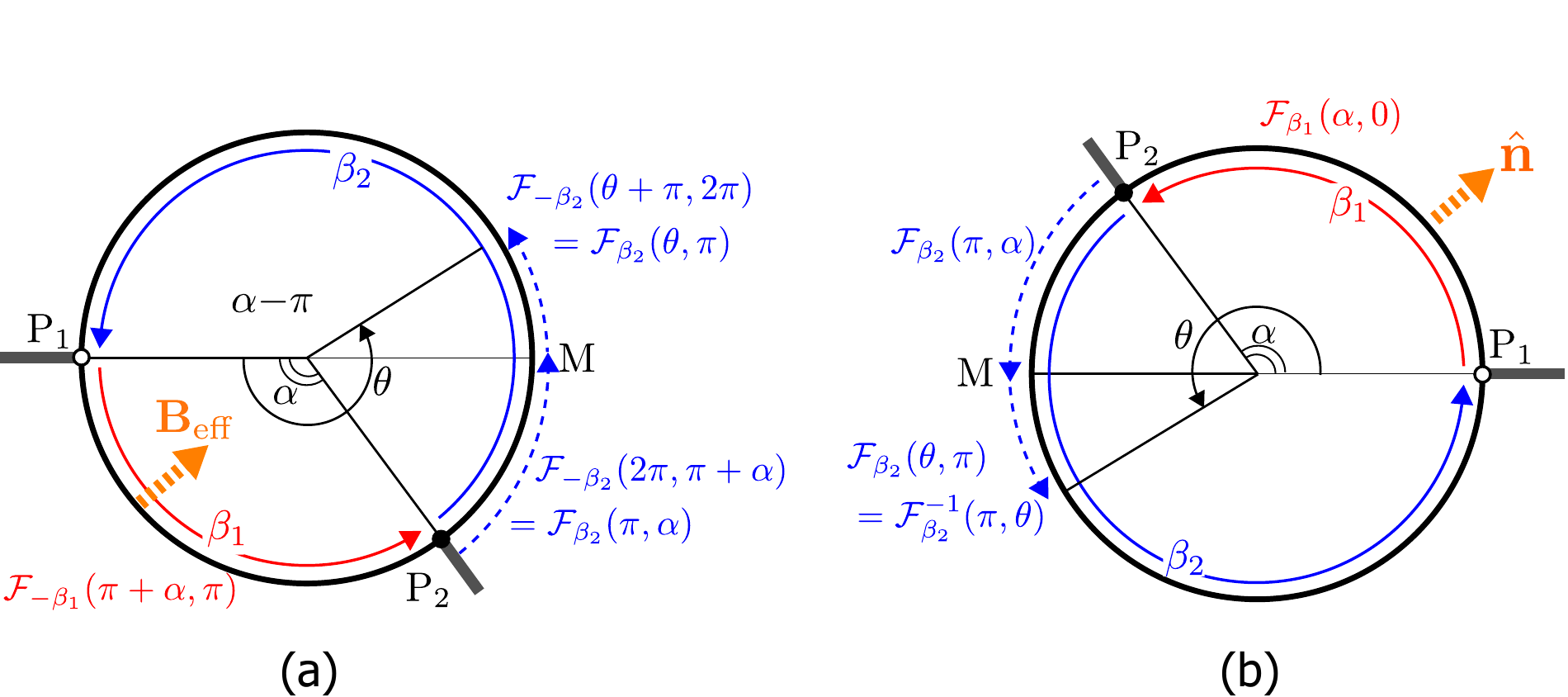}
\end{center}
\caption{Non-Abelian AC phase for a ring with two different spin-orbit couplings $\beta_1$ and $\beta_2$ can be calculated 
as a product of partial AC-phases: (a) the original ring interferometer (in the coordinate system used in 
Secs.~\ref{sec:Ring} and \ref{sec:Ringinhom}) and (b) the equivalent system used in the actual calculations.   
\label{fig:two-beta}}
\end{figure}
\end{widetext}
\section{Solving matching conditions}
\label{sec:solving-matching-cond}
In this appendix, we sketch the solutions to the matching conditions in Secs.~\ref{sec:Ring} and \ref{sec:Ringinhom}.   
The matching conditions \eqref{eqn:match_left} and  \eqref{eqn:Matchright} give a set of coupled equations for the six unknown matrices 
$A_+,A_-,B_+,B_-,r,t$. Let us write this set of equations in the matrix notation.  
(Each entry in the following matrices is a $2 \times 2$ matrix representing spin $\uparrow/\downarrow$):
\begin{equation}
M(k,\alpha;\beta_1,\beta_2) 
( A_+ , A_- , B_+ ,B_- , r ,t )^{\text{T}} 
 = (1,1,1,0,0,0)^{\text{T}}  
 \end{equation}
with $M$ given by
\begin{widetext}
\begin{equation}
M \equiv 
\begin{pmatrix}
F_{\text{AC}}(2 \pi; \beta)&F_{\text{AC}}(2 \pi; \beta)&0&0&-1&0 \\
 0&0&1&1&-1&0 \\
 F_{\text{AC}}(2 \pi; \beta)& -F_{\text{AC}}(2 \pi; \beta)&1&-1&1&0\\
 F_{\text{AC}}(\alpha;\beta) \be^{ik(2 \pi-\alpha)}& F_{\text{AC}}(\alpha;\beta) \be^{-ik(2 \pi-\alpha)}&0&0&0&-1\\
 0&0& F_{\text{AC}}(\alpha;\beta) \be^{ik\alpha}&F_{\text{AC}}(\alpha;\beta) \be^{-ik\alpha}&0&-1\\
 F_{\text{AC}}(\alpha;\beta) \be^{ik(2 \pi-\alpha)}& - F_{\text{AC}}(\alpha;\beta) \be^{-ik(2 \pi-\alpha)}&
 F_{\text{AC}}(\alpha;\beta) \be^{ik\alpha}&-F_{\text{AC}}(\alpha;\beta) \be^{-ik\alpha}&0&-1
 \end{pmatrix}
 \; .
 \label{Ring-Matrix}
 \end{equation}
Introducing a new set of variables
\begin{equation}
X_{\pm} \equiv A_{+} \pm A_{-} \; , \;\; Y_{\pm} \equiv B_{+} \pm B_{-} \; , \;\; r \; , \;\; 
\tilde{t} \equiv F_{\text{AC}}^{-1}(\alpha ;\beta )t = F_{\text{AC}}^{\dagger}(\alpha ;\beta )t \; ,
\end{equation}
we may recast the above equations into a new set of equations
\begin{equation}
\widetilde{M}
(X_{+} , X_{-} , Y_{+} , Y_{-} , r , \tilde{t} )^{\text{T}} 
= (1,1,1,0,0,0)^{\text{T}} 
\label{eqn:new-set-of-eqs}
\end{equation}
with a new coefficient matrix:
\begin{equation}
\widetilde{M} \equiv 
\begin{pmatrix}
F_{\text{AC}}(2 \pi ;\beta ) & 0 & 0 & 0 & -1 & 0 \\
 0 & 0 & 1 & 0 & -1 & 0 \\
 0 & F_{\text{AC}}(2 \pi ;\beta ) & 0 & 1 & 1 & 0 \\
 \cos \left[ (2 \pi - \alpha )k \right] & i \sin \left[ (2 \pi - \alpha )k \right] & 0 & 0 & 0 & - 1 \\
 0 & 0 & \cos (k \alpha ) & i \sin (k \alpha ) & 0 & - 1 \\
 i \sin \left[ (2 \pi - \alpha )k \right] & \cos \left[ (2 \pi - \alpha )k \right] & i \sin (k \alpha ) & \cos (k \alpha ) & 0 & - 1
\end{pmatrix}
\; .
\end{equation}
As the new set of equations \eqref{eqn:new-set-of-eqs} no longer contains $F_{\text{AC}}(\alpha ;\beta )$, 
we immediately see that the conductance $\text{Tr}(\tilde{t}^{\dagger}\tilde{t})=\text{Tr}(t^{\dagger}t)$ 
does not depend on $F_{\text{AC}}(\alpha ;\beta )$.  

It is relatively easy to eliminate $X_{\pm}$ and $\tilde{t}$ from these equations to obtain the following three equations for $X_{\pm}$ and $r$:
\begin{subequations}
\begin{align}
& F_{\text{AC}}(2 \pi ; \beta) X_{+} - r-1 =0  \;\; \leftrightharpoons \;\;  X_{+} = F_{\text{AC}}^{-1}(2 \pi ; \beta)(1+r)  
\label{eqn:matching-1} \\
\begin{split}
& \be^{-i (2 \pi -\alpha ) k} X_{+} 
+ \left\{ \cos (\alpha  k) F_{\text{AC}}(2\pi;\beta)  - \be^{-i (2 \pi -\alpha ) k}  \right\} X_{-} + r \be^{-i \alpha  k}  - \be^{i \alpha  k}  =0 
\label{eqn:matching-2}
\end{split}
\\
\begin{split}
& i \sin [(2 \pi -\alpha ) k] X_{+}
+ \left\{ \cos [(2 \pi -\alpha ) k] - \be^{-i \alpha  k} F_{\text{AC}}(2\pi ;\beta) \right\} X_{-} - 2 r \be^{-i \alpha  k}  =0 \; .
\label{eqn:matching-3}
\end{split}
\end{align}
\end{subequations}
If we plug $X_{+}$ [\eqref{eqn:matching-1}] into \eqref{eqn:matching-2} and \eqref{eqn:matching-3} to eliminate $X_{-}$,  
we obtain the matrix $r$:
\begin{equation}
\begin{split}
r =  & \left\{ F_{\text{AC}}(2\pi;\beta) + F^{-1}_{\text{AC}}(2\pi;\beta) 
+  \left[ \sin ((2 \pi -\alpha ) k) \sin (\alpha  k)-2 \be^{-2 i \pi  k} \right]  \right\}^{-1} \\
&  \left\{ -F_{\text{AC}}(2\pi;\beta) - F^{-1}_{\text{AC}}(2\pi;\beta) 
+ \left[ \sin ((2 \pi -\alpha ) k) \sin (\alpha  k)+2 \cos (2 \pi  k) \right]   \right\} \; .
\end{split}
\label{eqn:r-matrix-general}
\end{equation}  
Using the identity \eqref{eqn:identity-F-Finv}, we see that the right-hand side is in fact a scalar matrix 
given by Eq.~\eqref{eqn:r-matrix-ring}.   
\end{widetext}
\section{Solution of the square model by the Transfer Matrix Method}
\label{sec:solution-square}
Here we solve the scattering problem for the model whose Hamiltonian is introduced in Eq.~\eqref{H}, (see Fig.~\ref{SquareScatt}),  
and compute transmission and reflection amplitudes.  
Our aim is to solve the Schr\"odinger equation $H |\Psi \ra = \veps |\Psi \ra$ for the two component spinor $|\Psi \ra$, subject to 
scattering boundary conditions.  Here $\varepsilon=-2 \cos k$ is the scattering energy and $k$ is the wave number 
(where we have taken the lattice constant $a=1$).  For definiteness, we consider a scattering problem wherein an incoming electron 
approaches the link at $n=0$ from the left ($n<0$) in channel $\beta=1,2$ with spin direction $\mu=\pm=\ua,\da$.  
It can be reflected or transmitted into channel $\alpha = 1,2$ with spin direction $\sigma=\pm=\ua,\da$.  
Henceforth, the spinor wave functions and the scattering amplitudes depend on (and should carry) 
the initial quantum numbers $|\beta \mu \ra$. Thus, the corresponding reflection and transmission amplitudes 
are written as $r_{ \alpha \sigma;\beta \mu}$, and $t_{ \alpha \sigma;\beta \mu}$. 

We expand the spinor in a complete set of basis functions in the [chain$\otimes$site$\otimes$spin]  space.  
The basis functions are denoted by $|\alpha n  \sigma \ra$; explicitly, $|\alpha  n \ua \ra=|\alpha n \ra \otimes \binom{1}{0}$ 
and $|\alpha  n \da \ra = |\alpha n \ra \otimes \binom{0}{1}$. Thus,  
\begin{equation} \label{Psi}
| \Psi \ra_{\beta \mu} = \sum_{\alpha n \sigma } \psi_{\alpha \sigma;\beta \mu}(n) |\alpha n \sigma \ra~, \ \ \psi_{\alpha;\beta \mu}(n) = \begin{pmatrix} \psi_{\alpha \ua}(n) \\ \psi_{\alpha \da} (n) \end{pmatrix}_{\beta \mu}. 
\end{equation}

It is useful to use compact notation and define a $4$$\times$$4$ wave-function matrix 
$[{\bm \Psi}(n)]$ whose elements are the spinor components $\psi_{\alpha \sigma,\beta \mu}(n)$ 
defined in Eq.~(\ref{Psi}), 
\begin{equation} \label{psi4x4}
[{\bm \Psi}(n)]_{\alpha \sigma,\beta \mu}=\psi_{\alpha \sigma,\beta \mu}(n),
\end{equation}
where the order of rows (counting from the top) or columns (counting from the left) is
 $(1 \ua, 1 \da,2 \ua, 2 \da)$, equivalently, the 4 dimensional space is channel$\otimes$spin.  

Now we define the local $8$$\times$$8$ transfer matrices $T_n$, $n=-1,0,1,2$, and a total transfer matrix $T$,
\begin{equation} \label{Tn}
 \begin{pmatrix} {\bm \Psi}(n) \\ {\bm \Psi}(n-1) \end{pmatrix}=T_{n-1} \begin{pmatrix}{\bm \Psi}(n-1) \\ {\bm \Psi}(n-2) \end{pmatrix}, \quad  T=T_2T_{1}T_0T_{-1}
\end{equation}
The transfer matrices act on $8{\times}4$ wave function matrices.  The above construction implies that the total transfer matrix $T$ 
across the square satisfies
\begin{equation} \label{transfer}
\begin{pmatrix} {\bm \Psi}(3) \\ {\bm \Psi}(2) \end{pmatrix} =T  \begin{pmatrix}{\bm \Psi}(-1) \\ {\bm \Psi}(-2) \end{pmatrix}.
\end{equation}
Knowing the $8{\times}8$ transfer matrix $T$, one obtains the $4{\times}4$ transmission and reflection matrices $t$ and $r$ 
with elements $t_{ \alpha \sigma; \beta \mu}$ and $r_{ \alpha \sigma; \beta \mu}$.  
Starting from Eq.~\eqref{transfer} we find, 
\begin{equation} \label{psirt}
\begin{split}
& {\bm \Psi}(-1)=I_{4 \times 4}+r, \ \ {\bm \Psi}(-2)= \be^{-i k}I_{4 \times 4}+ \be^{i k} r, \\ 
&  {\bm \Psi}(2)=t, \ \  {\bm \Psi}(3)= \be^{i k} t .
\end{split}
\end{equation} 
This enables us to express $r$ and $t$ in terms of the four $4$$\times$$4$ blocks of $T$, denoted as $T_{ij}$, with $(i,j=1,2)$.  
The explicit expressions are:
\begin{equation}
\begin{split}
r= & [\be^{ik}(T_{21}-T_{12})+\be^{2 i k}T_{22}-T_{11}]^{-1}   \\
& \quad [T_{11}+\be^{-i k}T_{12}-\be^{ik}T_{21}-T_{22}]   \\
t= & \be^{ik}T_{11}(I_{4 \times 4}+r)+ \be^{-ik}T_{12}(e^{-i k}I_{4 \times 4}+ \be^{ik}r). 
\end{split}
\label{sol}
\end{equation}
As a test of the correctness of these relations one can confirm the unitarity and time-reversal constraints,
\begin{equation} \label{unitarity1}
\begin{split}
& {\mathrm{Tr}}[t^\dagger t+r^\dagger r]=4, \quad
t'_{\alpha \nu \beta \mu}=(-1)^{\nu-\beta}t^*_{\beta \bar{\mu }\alpha \bar{\nu}}, \\
& r_{\alpha \nu \beta \mu}=(-1)^{\nu-\mu}r_{\beta \bar{\mu } \alpha \bar{\nu}}, \ 
r'_{\alpha \nu \beta \mu}=(-1)^{\nu-\mu}r'_{\beta \bar{\mu } \alpha \bar{\nu}} \; .
\end{split}
\end{equation}
Here $\bar {\sigma}=-\sigma$, and $t'$ and $r'$ are the transmission and reflection matrices for scattering 
of incoming electrons from the right.  Note that these relations connect matrix elements of the transmission matrices 
{\em on different sides of the sample}, and matrix elements of the reflection matrices {\em on the same side of the sample}.   
These relations imply the absence of spin-flip in the reflection amplitude of the same channel, 
i.e., $r_{\alpha \nu \alpha \bar {\nu}}=0$, and also that the diagonal elements of the reflection matrix are equal, 
$r_{\alpha \ua, \alpha \ua}=r_{\alpha \da, \alpha \da}$ for each channel.  

It remains to determine the $8$$\times$$8$ local transfer matrices $\{ T_n \}$. A glance at Fig.~\ref{SquareScatt} suggests 
that there are kinds of sites: (1) For $n<0$ and $n>1$ the coordination number is 2 and the two links attached to it are bare. 
(2) For $n=0$ and $n=1$ the coordination number is 3, one link is bare and two links are ``dressed'' with SU(2) hopping matrices.  
This respectively requires two slightly different definitions of the local transfer matrices. For this purpose it is useful 
to define the following $4$$\times$$4$ matrices:
$$ 
X \equiv \begin{pmatrix} 0& \be^{i \beta_x \sigma_x} \\ \be^{-i \beta_x \sigma_x}&0 \end{pmatrix}, \ \ 
Z \equiv I_{2 \times 2} \otimes \be^{i \beta_z \sigma_z}. 
$$
After some algebra we find,
\begin{equation} \label{Tnexplicit} 
\begin{split}
& T_{-1}=
\begin{pmatrix} 
-\tfrac{\varepsilon}{t} & -1\\
1& 0 
\end{pmatrix}, \;  
T_0=\begin{pmatrix} -Z^\dagger(\tfrac{\varepsilon}{t}+X)&-Z\\
1& 0 \end{pmatrix}, \\ 
& T_1=
\begin{pmatrix} 
-Z^\dagger(\tfrac{\varepsilon}{t}+X) &  -Z^2\\
1& 0 
\end{pmatrix},   \;  
T_{2}=
\begin{pmatrix} 
-(\tfrac{\varepsilon}{t}+X)&-Z \\
1& 0 
\end{pmatrix} 
\; ,
\end{split}
\end{equation}
where every entry in these matrices is a $4$$\times$$4$ matrix in channel$\otimes$spin space.

\bibliographystyle{jpsj}
\bibliography{./references/mesoscopic,./references/Avishai}
\end{document}